# The Generalized Plasma Focus Problem and its Application to Space Propulsion

**S K H Auluck**
International Scientific Committee on Dense Magnetized Plasmas,
Hery 23, P.O. Box 49, 00-908 Warsaw, Poland, skhauluck@gmail.com

Abstract:

Space propulsion is unique among many proposed applications of the Dense Plasma Focus in being critically dependent on the availability of a scaling theory that is well-grounded in physics, in conformity with existing experimental knowledge and applicable to experimentally untested configurations. This paper derives such a first-principles-based scaling theory and illustrates its application to a novel space propulsion concept, where the plasma focus sheath is employed as a power density amplifying mechanism to transport electric energy from a capacitive storage to a current-driven fusion load. For this purpose, a Generalized Plasma Focus problem is introduced and formulated. It concerns a finite, axisymmetric plasma, driven through a neutral gas at supersonic speed over distances much larger than its typical gradient scale length by its azimuthal magnetic field while remaining connected with its pulse power source through suitable boundaries. The Gratton-Vargas equation is rederived from the scaling properties of the equations governing plasma dynamics and solved for algebraically defined initial (insulator) and boundary (anode) surfaces. Scaling relations for a new space propulsion concept are derived. This consists of a modified plasma focus with a tapered anode that transports current from a pulsed power source to a consumable portion of the anode in the form of a hypodermic needle tube continuously extruded along the axis of the device. When the tube is filled with deuterium, the device may serve as a small-scale version of magnetized liner inertial fusion (MAGLIF) that could avoid failure of neutron yield scaling in a conventional plasma focus.

I.  Introduction:

The dense plasma focus (DPF) [1,2,3] is well-known for being a prolific source of neutrons [4], fast ions [5], electrons [6] and soft x-rays [6] as well as a uniquely useful plasma environment [6]. Applications of the dense plasma focus in materials science [7], nanotechnology [8,9], biomedicine [10], production of short-lived radioisotopes for medical imaging [11] have been discussed in literature. Another potential application that has received attention is space propulsion.

The concept of plasma focus based space propulsion belongs to the more general class of electric propulsion concepts – variously known as Hall thrusters or Magneto-Plasma-Dynamic (MPD) thrusters – where the Hall effect is used to accelerate the propellent in plasma form to a high velocity using electric energy input. Plasma focus based micro-thrusters have been developed for steering and motion control applications on nanosatellites [12, 13] using fast ejection of plasma containing ablated anode and insulator material. Plasma focus based space propulsion has also been found to have interesting features [14-17] for interplanetary or deep space missions using its fusion output as reflected in the following quote from a study [16] that compares four types of high thrust propulsion systems: chemical (based on liquid hydrogen + liquid oxygen), nuclear thermal based on hydrogen heated using fission, inertial fusion using fusion reaction products to heat propellent and the DPF: "Comparison between the DPF and other existing and proposed systems show the DPF to be one of the few systems capable of combining high specific impulse with a large thrust. The thrust projected for the DPF rates it as one of the more powerful fusion propulsion systems." "The DPF propulsion system can have a wide range of values" for parameters suitable for the "[f]our main mission classes …: surface-to-orbit, inner solar system, outer solar system, and extra solar." "…the high Isp and thrust-to-weight ratio of the DPF make it a forerunner in space propulsion systems."

Unlike Hall / MPD thrusters, the DPF based propulsion concepts utilise electric energy input and the Hall effect to achieve amplification of energy using fusion reactions, where the unburnt fuel heated by fusion reaction products acts as the propellant providing enhanced thrust and greater power density as compared with non-nuclear Hall /MPD thrusters. Practical development of such concepts is predicated on achieving significant fusion gain and is still in the future. However, such practical development requires contemporary conceptual engineering studies [16,17] as an essential prerequisite.

A recent comprehensive review [3] of the scientific status of the DPF reveals that the fusion energy output of a plasma focus involves a complex process [3], not fully understood as yet, in which ions are accelerated and trapped in spontaneously generated magneto-plasma structures which make them repeatedly interact with a dense plasma target. Scaling failure of this process is experimentally observed but no theoretical understanding or empirical workaround exists as of now [3]. Any interplanetary space propulsion concept based on this process would thus be in jeopardy as far as the current state of knowledge about the plasma focus fusion process is concerned. Since Mars missions are being actively pursued by national space agencies of many countries and even some private commercial companies, it becomes a topic of current relevance to address this issue and to search for more feasible alternatives.

The apparent attractiveness of the plasma focus for space propulsion arises primarily from its demonstrated ability to amplify power density. It extracts electrical energy stored over a period of several tens of seconds in a capacitance spread over a volume of several cubic meters, into the plasma focus chamber with a volume of few litres on the scale of several (tens) microseconds and delivers it to a plasma occupying a few cubic millimetres in a few tens of nanoseconds. This amplification of power density is a consequence of the transport of energy by a moving plasma current sheath acting as a plasma flow switch. It may be possible to deliver this energy to another kind of current-driven fusion load. Two examples of such well-researched fusion load concepts are Z-pinch driven dynamic hohlraum [18,19] and the magnetized liner inertial fusion (MAGLIF) [20,21]. By combining the power density amplifying property of the plasma focus sheath with these or similar fusion loads, the attractive features of the DPF as a potential space propulsion concept would be retained without being plagued by uncertainties related to the fusion mechanism in the conventional Mather type plasma focus.

Unlike other applications of the DPF, space propulsion for interplanetary and deep space missions critically depends on the availability of a scaling theory that is well-grounded in physics, is in conformity with existing knowledgebase and can be credibly utilized for predicting power density amplification for newly conceived configurations over multiple scales of physical size and energy input. The multi-dimensional parameter space of a propulsion concept needs to be scanned for identifying options that have favourable benefits and costs without incurring an unviable expense in effort and time in their search. A scaling theory provides relatively short computational procedures to generate relevant ballpark numbers necessary to decide whether an option needs further examination or can be safely discarded without missing significant opportunities. It is also required for carrying over insights gained in laboratory experiments to conceptual engineering research [16,17] which must precede an engineering research and development effort.

The Dense Plasma Focus has been a subject of MHD and kinetic modelling studies for quite a long time [22-34] which have been reviewed comprehensively [3]. One feature of modern computational models [28,29] of the Dense Plasma Focus is that they give better agreement with experiments if the values of circuit inductance and resistance are suitably chosen to be different from measured values. Other simple models such as the Lee model [35] or the Resistive Gratton-Vargas (RGV) model [36-38] require an experimental current waveform from an existing facility in order to determine model parameters using an iterative fitting procedure. They also make simplifying assumptions that are contrary to reality. The Lee model neglects the existence and role of the plasma formation region, the delay between the start-of-current and plasma propagation and the curved shape of the plasma in the rundown and radial phases. The GV model [36-39] neglects the delay between the start-of-current and plasma propagation, the existence of front and rear boundaries of the plasma sheath and in fact, even the existence of a physical plasma. In both models, this neglect is compensated for by a fitting procedure applied to an experimental waveform from an existing facility. Their inability to link their predictions for newly conceived and experimentally untested configurations with first principles using a transparent logic makes them unsuitable for credible explorations of space propulsion concepts.

The present paper aims to derive a scaling theory that can generate ballpark estimates of power density amplification for such newly conceived and experimentally untested configurations from considerations rooted in first principles and also to illustrate its application to space propulsion.

The purpose of this paper is to introduce and formulate the Generalized Plasma Focus (GPF) problem that includes this scaling theory as one component. It concerns a finite, axisymmetric plasma driven through a neutral gas medium at supersonic speed over distances much larger than its typical gradient scale length by its azimuthal magnetic field while remaining connected with its pulse power source through suitable boundaries.

It is shown that equations governing plasma dynamics can be rendered into a dimensionless form using scaling parameters that characterize the physical plasma and its relation with the pulse power source. The dimensionless equations then become independent of the plasma device except for a single parameter that depends on the dimensionless time derivative of the logarithm of the dimensionless current profile. Therefore, they universally describe the physics of all devices that fall within the ambit of the GPF.

The Gratton-Vargas (GV) equation [36-40] is rederived under very general assumptions. It describes evolution of an imaginary 3-dimensional surface of revolution whose normal velocity equals the scaling parameter for velocity. The evolution of all scaling parameters in real space and time is determined from the solutions of the Gratton-Vargas equation with algebraically defined initial and boundary conditions. This includes calculation of the temporal profile of current. The delay between the start of current and plasma propagation is explicitly addressed. The result is a scaling theory, well-grounded in physics, that can facilitate preliminary evaluation of any newly conceived and experimentally untested space propulsion concept that conforms to the definition of the GPF. Predictions of this scaling theory are shown to compare favourably with experimental knowledgebase.

Utility of this scaling theory is illustrated with reference to a novel space propulsion concept that can be studied as a laboratory experiment for validating the scaling theory and perhaps also as a model of a non-nuclear electric propulsion engine for near earth missions. It can also be configured as a laboratory neutron source that may circumvent the neutron yield scaling failure [3] in the conventional Mather type plasma focus.

In this concept, a plasma-focus-like device with a profiled anode acts as a plasma flow switch to transport the current from a pulsed power source to the "fuel". The "fuel" is a consumable portion of the anode in the form of a metal wire (or a hypodermic needle tube) continuously extruded along its axis. The plasma flowing over the surface of this fuel would launch a travelling ablation wave accompanied by a magnetic pressure wave. This would convert the metal wire (or tube) into a plasma that is radially confined and free to flow only along the axis at its Alfven velocity. The mass of the wire plasma times its Alfven velocity provides an estimate of the impulse per shot for a laboratory model of a non-nuclear electric propulsion engine. Its measurement and scaling behaviour as a function of device and energy storage parameters can be used to benchmark the scaling theory. The thrust is governed by the shot repetition rate. Along with the evaluation of the energy storage requirements, this example can form the basis for scanning the parameter space for optimizing the thrust to weight ratio. Conical collapse of a hollow tube (instead of a solid wire) by the travelling ablation wave can produce additional power density amplification by the shaped charge effect discussed later, that might be of interest for a non-nuclear electric propulsion engine. If the tube is filled with deuterium gas and perhaps also immersed in an axial magnetic field, it can serve as a laboratory fusion source that can be studied as a surrogate for a potential aneutronic fusion drive. Much of the fusion physics modelling for the z-pinch driven dynamic hohlraum [18, 19] and magnetized liner inertial fusion (MAGLIF) [20, 21] concepts can be extrapolated to this configuration as well. This fusion source is likely to circumvent the neutron yield scaling failure observed [3] in a conventional plasma focus by making available a compact high density fusion capsule that can utilise the energy transported by the plasma flow switch in spite of the failure of an ionization stability condition discussed later.

The scope of this paper is limited to the presentation of the Generalized Plasma Focus problem, its comparison with experimental knowledgebase and demonstration of its usefulness as a scaling theory applicable for space propulsion. The concept of space propulsion is described mainly as an illustration of the kind of problem that could be addressed using this scaling theory. This paper does not intend to make any claims regarding the feasibility of the proposed space propulsion concept or its relative merit in comparison to its peer concepts.

This paper is organized as follows: Section II defines the Generalized Plasma Focus problem, derives the Gratton-Vargas equation, establishes the procedure for applying initial and boundary conditions using algebraically defined initial (insulator) and boundary (anode) surfaces and sets up and solves the circuit equation resulting in model closure. Section III compares some examples of theoretical results with experimental knowledgebase. Section IV applies the formalism to an illustrative space propulsion problem where a profiled anode acts as a plasma flow switch to rapidly transfer energy from a pulsed power source to a metal wire that forms a consumable portion of the anode. Section V discusses some aspects of laboratory experiments for further study of the proposed space propulsion concept. Section VI concludes the paper with a summary and discussion.

II. <u>The Generalized Plasma Focus (GPF) problem</u>.

This section is a condensed summary of a more detailed tutorial treatment published elsewhere [40].

A. The physical basis of the Generalized Plasma Focus problem:

The Generalized Plasma Focus problem consists of the following elements

(a) An axis of symmetry, taken to be the z-axis of a cylindrical coordinate system fixed with the device.

(b) A uniform gaseous medium of mass density $\rho_0$.

(c) A scale length 'a' that represents the physical size of the problem.

(d) A plasma region within the gaseous medium, obeying axial symmetry, that carries a current density distribution $\vec{J}(\vec{r},t)$ with a total current $I(t)$.

(e) A well-characterized pulse power source which supplies the current $I(t)$

(f) A conducting boundary that keeps the plasma region electrically connected to the power source.

(g) The plasma is assumed to be free to move while maintaining electrical contact with the conducting boundary over distances larger than its typical gradient scale length.

(h) The associated magnetic field distribution is such that its toroidal component $\vec{B}_T \equiv B_\theta(\vec{r},t)\hat{\theta}$ is much larger in magnitude as compared with its poloidal component $\vec{B}_P \equiv B_r(\vec{r},t)\hat{r} + B_z(\vec{r},t)\hat{z}$ everywhere and at all times, so that it can be neglected in the first approximation.

The single-fluid MHD model is assumed to be applicable in this paper, but it need not form part of the specification of the GPF, where more detailed models that account for presence of variable concentrations of ionized and neutral species may be invoked. Given these elements, the GPF considers the question of transforming the governing equations of plasma dynamics (which could be any variety of single or multi-species MHD fluid model) into a dimensionless form using the elements defined above. In the following discussion, the physical parameters of the plasma are expressed as products of a scaling factor, denoted by the subscript 0 (except for the scale length 'a') and a scaled parameter denoted by an overtilde: mass density $\rho \equiv \rho_0 \tilde{\rho}$; radial and axial coordinates $r \equiv a\tilde{r}$; $z \equiv a\tilde{z}$; gradient operator $\vec{\nabla} \equiv a^{-1}\tilde{\nabla}$; velocity $\vec{v} = v_0 \tilde{v}$; magnetic field $\vec{B} \equiv B_0 \tilde{B}$, electric field $\vec{E} = E_0 \tilde{E}$, pressure $p = p_0 \tilde{p}$, current $I(t) = I_0 \tilde{I}(\tau)$. The non-dimensional time-equivalent parameter $\tau$ is defined later.

The scale factor for the magnetic field is chosen to be the magnetic field produced by a straight current-carrying conductor at a radius r:

$$B_0 = \frac{\mu_0 I(t)}{2\pi a \tilde{r}} \tag{1}$$

The rationale for this choice is the fact that the plasma propagates over distances much larger than its typical gradient scale length according to element (g) of the GPF specification. Behind the gradient-bearing region of the plasma, the magnetic field is given by relation (1). This choice is the core of the physical basis of the scaling theory.

Then it is easily verified that the single-fluid equation of motion of the plasma

$$\rho\left(\frac{\partial \vec{v}}{\partial t} + (\vec{v}\cdot\vec{\nabla})\vec{v}\right) = \vec{J}\times\vec{B} - \vec{\nabla}p \tag{2}$$

can be expressed in a non-dimensional form by choosing

$$v_0 \equiv \frac{B_0}{\sqrt{2\mu_0\rho_0}} ; \quad p_0 \equiv \frac{B_0^2}{2\mu_0} ; \tag{3}$$

$$\frac{d\tau}{dt} = \frac{I(t)}{Q_m} \Rightarrow \frac{\partial}{\partial t} \equiv \frac{d\tau}{dt}\frac{\partial}{\partial \tau} = \frac{I(t)}{Q_m}\frac{\partial}{\partial \tau} ; \quad Q_m = \pi\mu_0^{-1}a^2\sqrt{2\mu_0\rho_0} \tag{4}$$

The resulting dimensionless equation

$$\tilde{\rho}\left\{\frac{\partial \tilde{v}}{\partial \tau} + \tilde{v}\frac{\partial}{\partial \tau}\text{Log}(\tilde{I}(\tau)) + \frac{1}{2}\tilde{v}\cdot\tilde{\nabla}(\tilde{r}^{-1}\tilde{v})\right\} = -\tilde{B}\times\left(\tilde{\nabla}\times(\tilde{r}^{-1}\tilde{B})\right) - \frac{1}{2}\tilde{r}\tilde{\nabla}(\tilde{r}^{-2}\tilde{p}) \tag{5}$$

is connected with the GPF only through the term proportional to $\partial_\tau \text{Log}(\tilde{I}(\tau))$. Other governing equations of plasma dynamics are also found to be rendered in a dimensionless form through the same choice of scaling factors and they are also related to the GPF through a similar term.

The plasma dynamics can therefore be resolved into two subproblems:

(1) determination of the scaling factors in the physical space and time;

(2) determination of the scaled quantities in the dimensionless space and time.

The two subproblems are weakly interdependent and can be separately handled iteratively, neglecting this weak interdependence in the first iteration. This aspect is seen in Section II(D) below.

The first subproblem is the scaling theory, which is formulated in this paper in terms of the Gratton-Vargas (GV) equation [36-40] for an imaginary surface $\psi(\tau,\tilde{r},\tilde{z}) = 0$, called the reference GV surface, whose normal velocity equals the scaling velocity and which complies with applicable initial and boundary conditions.

The second subproblem constitutes a physical theory that expresses the temporal evolution of the spatial distribution of scaled plasma parameters in dimensionless space and time. It includes a variety of physical processes which affect concentrations of ionized and neutral species that enter the specifications of pressure and internal energy as functions of density. The spatial distributions of scaled plasma parameters such as density, velocity, pressure and magnetic field must move along with the imaginary reference GV surface $\psi(\tau,\tilde{r},\tilde{z}) = 0$ by definition. They could then be approximated as functions of $\psi(\tau,\tilde{r},\tilde{z})$ which can be used as a similarity variable to reduce the system of partial differential equations governing plasma parameters into a system of ordinary differential equations. The absence of all information of the physical device from the equations governing evolution of scaled plasma variables implies that they are universally applicable to all configurations which conform to the definition of the GPF and the physics model invoked. This program is outside the scope of the present paper and would be pursued separately. However, the universal applicability of the equations governing evolution of scaled plasma parameters implies the existence of scaling laws that must be applicable to a whole class of device configurations irrespective of their physical size or the scale of energy storage. This has in fact been observed in the case of the classical plasma focus and is discussed in more detail in Section III.

B. The Gratton-Vargas equation:

Define an imaginary surface of revolution $\psi(t,r,z) = 0$ having an instantaneous local normal velocity $v_n = v_0$. By definition, the convective derivative of $\psi$ is zero:

$$\partial_t \psi + \vec{v} \cdot \vec{\nabla} \psi = 0 \tag{6}$$

Therefore,

$$v_n = \vec{v} \cdot \left( \vec{\nabla} \psi / \left| \vec{\nabla} \psi \right| \right) = -\partial_t \psi \bigg/ \sqrt{\left(\partial_r \psi\right)^2 + \left(\partial_z \psi\right)^2} \tag{7}$$

From (1), (4), and (7), one gets the Gratton-Vargas (GV) equation

$$\partial_\tau \psi + \sqrt{\left(\partial_{\tilde{r}} \psi\right)^2 + \left(\partial_{\tilde{z}} \psi\right)^2} \frac{1}{2\tilde{r}} = 0 \tag{8}$$

It is seen to have infinitely many solutions of the form $\psi = c_1 \psi_1 + \psi_0$ where $\psi_1$ is one of its solutions and $c_1$ and $\psi_0$ are constants. So, the solutions of the GV equations are defined to within an arbitrary multiplicative constant, which can be chosen to provide a convenient normalization.

The additive constant $\psi_0$ generates a family of integral surfaces of the Gratton-Vargas equation. A particular subset of integral surfaces, labelled by $\tau$, which satisfy the given initial and boundary conditions and are of the form $\psi(\tau,\tilde{r},\tilde{z}) = 0$, is designated as the set of reference GV surfaces (RGS), which represents the idealized location of the interface between current-free and current-bearing regions of the plasma in the scaling theory. Integral surfaces which satisfy the initial and boundary conditions and which are of the form $\psi(\tau,\tilde{r},\tilde{z}) = \psi_0 \neq 0$ are designated as associated GV surfaces (AGS) for a given reference GV surface. When scaled plasma parameters such as density, velocity, pressure and magnetic field are expressed as functions of $\psi(\tau,\tilde{r},\tilde{z})$ as a similarity parameter, their spatial distribution spans the spatial continuum defined by the associated GV surfaces for both positive and negative values of $\psi_0$ for a given RGS.

Detailed tutorial discussions of solutions of this equation are found elsewhere [36-40]. Some relevant results are summarized below along with some new additions.

The GV equation (8) can be cast in the Jacobi form [42] $p_\tau + H = 0$ where $p_\tau \equiv \partial_\tau \psi$, $p_{\tilde{r}} \equiv \partial_{\tilde{r}} \psi$, $p_{\tilde{z}} \equiv \partial_{\tilde{z}} \psi$, and $H = (2\tilde{r})^{-1} \sqrt{p_{\tilde{r}}^2 + p_{\tilde{z}}^2}$.

The associated Hamilton-Jacobi equations for the characteristic line elements (CLE) [42] are

$$\frac{d\tilde{r}}{d\tau} = \frac{\partial H}{\partial p_{\tilde{r}}} = \frac{p_{\tilde{r}}}{2\tilde{r}\sqrt{p_{\tilde{r}}^2 + p_{\tilde{z}}^2}} = \frac{p_{\tilde{r}}}{4\tilde{r}^2 H} \qquad (9)$$

$$\frac{d\tilde{z}}{d\tau} = \frac{\partial H}{\partial p_{\tilde{z}}} = \frac{p_{\tilde{z}}}{2\tilde{r}\sqrt{p_{\tilde{r}}^2 + p_{\tilde{z}}^2}} = \frac{p_{\tilde{z}}}{4\tilde{r}^2 H} \qquad (10)$$

$$\frac{dp_{\tilde{r}}}{d\tau} = -\frac{\partial H}{\partial \tilde{r}} = \sqrt{p_{\tilde{r}}^2 + p_{\tilde{z}}^2} \frac{1}{2\tilde{r}^2} = \frac{H}{\tilde{r}} \qquad (11)$$

$$\frac{dp_{\tilde{z}}}{d\tau} = -\frac{\partial H}{\partial \tilde{z}} = 0 \qquad (12)$$

For equations of the Jacobi form [42], a Hamiltonian that is not explicitly dependent on time is always conserved:

$$\frac{dH}{d\tau} = \frac{\partial H}{\partial \tau} + \frac{\partial H}{\partial p_{\tilde{r}}}\frac{dp_{\tilde{r}}}{d\tau} + \frac{\partial H}{\partial p_{\tilde{z}}}\frac{dp_{\tilde{z}}}{d\tau} + \frac{\partial H}{\partial \tilde{r}}\frac{d\tilde{r}}{d\tau} + \frac{\partial H}{\partial \tilde{z}}\frac{d\tilde{z}}{d\tau}$$
$$= \frac{\partial H}{\partial \tau} - \frac{\partial H}{\partial \tilde{r}}\frac{\partial H}{\partial p_{\tilde{r}}} - \frac{\partial H}{\partial \tilde{z}}\frac{\partial H}{\partial p_{\tilde{z}}} + \frac{\partial H}{\partial \tilde{r}}\frac{\partial H}{\partial p_{\tilde{r}}} + \frac{\partial H}{\partial \tilde{z}}\frac{\partial H}{\partial p_{\tilde{z}}} = \frac{\partial H}{\partial \tau} = 0 \tag{13}$$

Since the GV equation (8) is unchanged by multiplying its solution with an arbitrary constant, this constant can be chosen so as to normalize the universal invariant H to unity. In subsequent discussions, the symbol H for the Hamiltonian is used as a placeholder, with the understanding that its numerical value is normalized to unity.

Equation (12) shows that $p_{\tilde{z}} \equiv \partial_{\tilde{z}}\psi$ is constant along the CLE. The two invariants are used [36-40] to define a new invariant, constant along the CLE, given by

$$N \equiv \frac{p_{\tilde{z}}}{2H} \tag{14}$$

which gives

$$p_{\tilde{r}} = sp_{\tilde{z}}\sqrt{(\tilde{r}/N)^2 - 1} = 2sH\sqrt{\tilde{r}^2 - N^2} \; ; s = \mathrm{sign}(d\tilde{r}/d\tau) \tag{15}$$

Then equations (9) and (10) can be written as

$$\frac{d\tilde{r}}{d\tau} = \frac{s\sqrt{\tilde{r}^2 - N^2}}{2\tilde{r}^2} \tag{16}$$

$$\frac{d\tilde{z}}{d\tau} = \frac{N}{2\tilde{r}^2} \tag{17}$$

From (16) and (17)

$$\frac{d\tilde{z}}{d\tilde{r}} = \frac{d\tilde{z}}{d\tau}\bigg/\frac{d\tilde{r}}{d\tau} = s\frac{N}{\sqrt{\tilde{r}^2 - N^2}} \tag{18}$$

Integration of (18) for $N \neq 0$ gives the following family C of curves of constant N, which are perpendicular to the GV surface.

$$\frac{\tilde{z}}{N} - s\mathrm{ArcCosh}\left(\frac{\tilde{r}}{|N|}\right) = C_1 = \text{constant}$$

(19)

Integration of (16) gives the location of the integral surface on the family C of curves at the dimensionless time $\tau$.

$$\frac{\tilde{r}}{|N|}\sqrt{\frac{\tilde{r}^2}{N^2}-1}+\text{ArcCosh}\left(\frac{\tilde{r}}{|N|}\right)-\frac{s\tau}{N^2}=C_2=\text{constant} \tag{20}$$

For the case of N=0, (16) shows

$$\tilde{r}^2 = s\tau + C_3, \tag{21}$$

The constants of integration $C_1, C_2, C_3$ are to be determined by the specifications of the problem in terms of initial and boundary conditions, as shown in the next subsection.

The function $\psi(\tau,\tilde{r},\tilde{z})$ can be calculated by integration of its total differential (22) along a CLE

$$d\psi = p_\tau d\tau + p_{\tilde{r}} d\tilde{r} + p_{\tilde{z}} d\tilde{z} \tag{22}$$

where, the preceding equations can be used to show that

$$p_{\tilde{z}} = 2HN = \text{constant} \tag{23}$$

$$p_{\tilde{r}} = 2Hs\sqrt{\tilde{r}^2 - N^2} \tag{24}$$

$$p_\tau = -H \tag{25}$$

This gives the functional form of $\psi(\tau,\tilde{r},\tilde{z})$ including a constant of integration $\psi_{GV}$ as

$$\psi(\tau,\tilde{r},\tilde{z}) = H\left(2N\tilde{z} + sN^2\left(\frac{\tilde{r}}{N}\sqrt{\left(\frac{\tilde{r}}{N}\right)^2 - 1} - \text{Arc Cosh}\left(\frac{\tilde{r}}{N}\right)\right) - \tau\right) - \psi_{GV} = \psi_0 \tag{26}$$

Substituting (19) and (20) in (26) and using the definition of the reference GV surface,

$$\psi_{GV} = HN^2(2C_1 + sC_2) \tag{27}$$

With H normalized to unity, (26) demonstrates that a change in $\psi_0$ is equivalent to a translation in $\tau$.

    C. Initial and boundary conditions.

The time-equivalent parameter $\tau$ in the Gratton-Vargas equation labels the members of the family of integral surfaces of the GV equation, called reference GV surfaces, which are curves in the $(\tilde{r},\tilde{z})$ plane. They greatly resemble the plasma focus sheath [41] and are taken to represent its shape and location during its propagation. The reference GV surfaces collectively span the entire space that can be visited by the plasma focus sheath. Since the typical gradient scale length, such as the width of the current carrying region, is assumed to be much smaller than the physical extent of this space, the reference GV surfaces $\psi(\tau,\tilde{r},\tilde{z})=0$ are assumed to approximate the path of the current flowing in the sheath as a function of $\tau$ within the context of the scaling theory.

The GPF problem specifies the initial conditions for the GV equation in terms of an initial surface, which would be the surface of the insulator in a conventional plasma focus. The boundary conditions are specified in terms of an arbitrary anode profile (with some restrictions which are discussed elsewhere [40]) and a straight cylindrical cathode.

A coordinate $\tilde{Z}$ is introduced for specifying the initial and boundary surfaces, measured in units of scale length 'a' along the axis of symmetry starting from the base of the anode. The initial surface is represented as a piecewise continuous curve in $(\tilde{r},\tilde{z})$ space:

$$\tilde{R}_I(\tilde{Z}) = \tilde{R}_I^{(i)}(\tilde{Z}), \quad \tilde{Z}_I^{(i-1)} \leq \tilde{Z} \leq \tilde{Z}_I^{(i)}, i=1,2\cdots,i_n; \quad \tilde{R}_I^{(i)}(\tilde{Z}_I^{(i)}) = \tilde{R}_I^{(i-1)}(\tilde{Z}_I^{(i)}); \qquad (28)$$

The anode profile is represented by another piecewise continuous curve with an external branch and an internal branch, the latter representing a cavity in the anode.

$$\begin{aligned}\tilde{R}_A(\tilde{Z}) &= \tilde{R}_{A,\text{ext}}^{(i)}(\tilde{Z}), \quad \tilde{Z}_A^{(i-1)} \leq \tilde{Z} \leq \tilde{Z}_A^{(i)}, i=1,2\cdots,i_m; \quad \tilde{R}_{A,\text{ext}}^{(i)}(\tilde{Z}_A^{(i)}) = \tilde{R}_{A,\text{ext}}^{(i-1)}(\tilde{Z}_A^{(i)}); \\ &= \tilde{R}_{A,\text{int}}^{(j)}(\tilde{Z}), \quad \tilde{Z}_A^{(i_m+j-1)} \geq \tilde{Z} \geq \tilde{Z}_A^{(i_m+j)}, j=1,2\cdots,j_m; \quad \tilde{R}_{A,\text{int}}^{(j)}(\tilde{Z}_A^{(i_m+j)}) = \tilde{R}_{A,\text{int}}^{(j+1)}(\tilde{Z}_A^{(i_m+j)});\end{aligned} \qquad (29)$$

The coordinate $\tilde{Z}$ increases monotonically in the external branch and decreases monotonically in the internal branch. The anode profile segment functions $\tilde{R}_{A,\text{ext}}^{(i)}(\tilde{Z})$ and $\tilde{R}_{A,\text{int}}^{(j)}(\tilde{Z})$ will not henceforth be distinguished in the text unless the context requires such distinction and both will be referred as $\tilde{R}_A^{(i)}(\tilde{Z})$. The points $(\tilde{R}_A^{(i)}(\tilde{Z}_A^{(i)}),\tilde{Z}_A^{(i)})$ etc. shall be referred as vertices. The following convention is also adopted: $\tilde{Z}_A^{(0)} \equiv \tilde{z}_I$, the normalized length of the insulator, and $\tilde{Z}_I^{(0)} = 0$. The cathode is taken to be a hollow cylinder with inner normalized radius $\tilde{r}_C$ connected to a metal base plate lying in the $\tilde{Z} = 0$ plane. The normalized height of the anode is denoted by $\tilde{z}_A \equiv \tilde{Z}_A^{(i_m)}$.

The GPF initial and boundary conditions are defined below

1. At $\tau = 0$, the initial shape of the solution corresponds to the initial surface,
$$\psi(\tau = 0, \tilde{r}, \tilde{z}) = \psi(\tau = 0, \tilde{R}_I(\tilde{Z}), \tilde{Z})$$

2. At every value of $\tau > 0$, the solution has a curve of intersection with the anode; i.e., $\psi(\tau, \tilde{R}_A(\tilde{Z}), \tilde{Z})$ is part of the solution of the equation at $\tau$.

3. It is assumed that the current that crosses over from the anode to the plasma does so *at or behind the reference GV surface*. This assumption more particularly specifies the reference GV surface to represent the interface between the current-free and current-bearing regions. That means that the current density at the anode surface has a zero in-surface component at the junction between the GV surface and the anode. The condition of continuity of current density at the junction implies that the reference GV surface must be perpendicular to the anode at the junction. This condition applies only to the forward motion of the GV surface and not to the reflected motion that begins after the plasma reaches the axis. The logic behind this argument may fail when the plasma displacement becomes comparable with its gradient scale length. An example is the stagnation of plasma when it reaches the axis in a conventional plasma focus.

4. The solution must join the anode profile with the cathode.

The first supplementary condition mentioned above implies that the characteristic should be perpendicular to the given initial surface profile at its intersection. From (18),

$$\frac{d\tilde{R}_I(\tilde{Z})}{d\tilde{Z}} = -\frac{d\tilde{z}}{d\tilde{r}} = -s\frac{N_I(\tilde{Z})}{\sqrt{\tilde{R}_I^2(\tilde{Z}) - N_I^2(\tilde{Z})}} \tag{30}$$

Therefore, in the initial plasma formation region

$$N_I(\tilde{Z}) = \frac{\tilde{R}_I(\tilde{Z})\left|d\tilde{R}_I(\tilde{Z})/d\tilde{Z}\right|}{\sqrt{\left(d\tilde{R}_I(\tilde{Z})/d\tilde{Z}\right)^2 + 1}} \tag{31}$$

Similarly, the third supplementary condition implies that the characteristic should be tangent to the anode profile at its intersection. Using (18)

$$\frac{d\tilde{r}}{d\tilde{z}} = S\frac{\sqrt{\tilde{r}^2 - N_A^{\ 2}(\tilde{Z})}}{N_A(\tilde{Z})} = \frac{d\tilde{R}_A(\tilde{Z})}{d\tilde{Z}} \tag{32}$$

This gives

$$N_A(\tilde{Z}) = S\frac{\tilde{R}_A(\tilde{Z})}{\sqrt{1 + \left(d\tilde{R}_A(\tilde{Z})/d\tilde{Z}\right)^2}}; S = \pm 1 \tag{33}$$

The second supplementary condition implies that the point of intersection between the GV surface and anode profile in the $(\tilde{r},\tilde{z})$ plane must move along the anode profile as a function of $\tau$ as described by (16), where $\tilde{r} = \tilde{R}_A(\tilde{Z})$ and $\tilde{z} = \tilde{Z}$:

$$\frac{d\tilde{Z}}{d\tau} = \frac{N_A(\tilde{Z})}{2\tilde{R}_A^{\ 2}(\tilde{Z})} \tag{34}$$

From (34), it is clear that the plus sign must be chosen in (33) when the solution is expected to begin at $\tilde{Z} = 0$ and move towards higher $\tilde{Z}$ as in the case of the external branch of the anode profile. There may be situations such as a hollow anode or a cavity in the anode face (see Ref 40 for examples), (represented by the internal branch in anode profile function (29)) when the solution is expected to move within the cavity towards decreasing values of $\tilde{Z}$. In such case, the negative sign would apply in (33). One could then set $S = \text{Sign}\left[d\tilde{Z}/d\tau\right]$.

Integration of (34) using (33) gives the value of $\tau(\tilde{Z})$ when the GV surface reaches the point $P_A(\tilde{R}_A(\tilde{Z}),\tilde{Z})$ on the external branch of anode profile

$$\tau(\tilde{Z}) - \tau_{i-1} = 2S \int_{\tilde{Z}_A^{(i-1)}}^{\tilde{Z}} \sqrt{1 + \left(d\tilde{R}_A^{(i)}/d\tilde{Z}\right)^2}\,\tilde{R}_A^{(i)}(\tilde{Z})d\tilde{Z}; \quad \tilde{Z}_A^{(i-1)} \leq \tilde{Z}(\tau) \leq \tilde{Z}_A^{(i)}; \tau_0 = 0 \tag{35}$$

The dimensionless time $\tau_i$ at which the GV surface reaches the vertex $\tilde{Z}_A^{(i)}$ is given by the recursion equations

$$\tau_i - \tau_{i-1} = 2S \int_{\tilde{Z}_A^{(i-1)}}^{\tilde{Z}_A^{(i)}} \sqrt{1 + \left(d\tilde{R}_A^{(i)}/d\tilde{Z}\right)^2}\,\tilde{R}_A^{(i)}(\tilde{Z})d\tilde{Z}; \tag{36}$$

where $\tau_i - \tau_{i-1}$ is the $i^{th}$ time zone that corresponds to the $i^{th}$ segment of the anode profile. The function $\tau(\tilde{Z})$ may be numerically inverted to obtain the coordinates $(\tilde{R}_A(\tilde{Z}(\tau)), \tilde{Z}(\tau))$ of the intersection of GV surface with the external anode surface for a given $\tau$.

The constants $C_1$ and $C_2$ in (19) and (20) are evaluated by using the values of $N_I(\tilde{Z})$ determined from (31) for the initial surface (28) or values of $N_A(\tilde{Z}(\tau))$ determined from (33) for the anode profile (29). The points $(\tilde{r}_{GV}, \tilde{z}_{GV})$ which obey both equations (19) and (20) can be found by parametrically defining [36-40]

$$\tilde{r}_{GV}(\alpha, N) = |N|\text{Cosh}(\alpha/2);$$
$$\tilde{z}_{GV}(\alpha, N, s) = NC_1(N, \tilde{Z}) + sN\alpha/2 \tag{37}$$

where $\alpha$ is the solution of the transcendental equation

$$F(\alpha) \equiv \text{Sinh}(\alpha) + \alpha = 2\left(\frac{\tilde{R}_A(\tilde{Z})}{|N(\tilde{Z})|}\sqrt{\frac{\tilde{R}_A(\tilde{Z})^2}{N^2(\tilde{Z})}-1} + \text{ArcCosh}\left(\frac{\tilde{R}_A(\tilde{Z})}{|N(\tilde{Z})|}\right) + s\frac{(\tau - \tau(\tilde{Z}))}{N^2(\tilde{Z})}\right) \tag{38}$$

At the vertices, the functions $N(\tilde{Z})$ given by (31) or (33) may be discontinuous for non-smooth profile functions. The GV surfaces are then constructed using (19) and (33) with values of N that continuously bridge the discontinuity. The procedure has been described in detail and illustrated with examples in a tutorial paper [40].

The case N=0 occurs in a Mather type device in the insulator region, with s=1. A practical anode profile will always have a flat top, however small its physical radius may be. On the flat top, the radius changes without any change in the axial coordinate indicating $d\tilde{R}_A(\tilde{Z})/d\tilde{Z} \to \infty$ so that in (33), $N_A(\tilde{Z}) \to 0$ near the centre of the anode. At first glance, equations (37) and (38) seem to suggest an unphysical divergent condition. However, it can be shown [40] that in this limit these equations lead to the quite physical and profoundly non-trivial result (which can also be derived from (21))

$$\tilde{r}^2 \to \left|\tilde{r}_{last}^2 + s(\tau - \tau_{last})\right| \text{ as } N_A(\tilde{Z}) \to 0 \tag{39}$$

where $\tilde{r}_{last}$ is the radius of the last vertex of the anode profile and $\tau_{last}$ is the time when that vertex is reached. The GV surface reaches the centre of the anode at $\tau_p = \tau_{last} + \tilde{r}_{last}^2$ when the radial velocity is directed towards the axis, i.e., for s=-1. It also shows that for $\tau > \tau_p$, the GV surface has a reflection, i.e., it starts moving outwards.

The construction of the reflected solution is discussed in detail elsewhere [39, 40] and will not be repeated here. Its main finding is that for $\tau > \tau_p$, the GV surface has three branches that intersect in a point: an Axially Expanding Front (AEF), a Radially Expanding Front (REF) and a connector branch (CN) which connects the intersection of the AEF and REF with the cathode. The AEF position on the axis is found [40] to obey a power law for a Mather type device

$$\tilde{z}_{APF} = 1.70246(\tau - \tau_p)^{0.443703} \tag{40}$$

The above discussion allows calculation of the constant of integration $\psi_{GV}$ in (27) in terms of the anode profile.

$$\psi_{GV} = 2N(\tilde{Z})\tilde{Z} - sN^2 \text{ArcCosh}\left(\frac{\tilde{R}_A(\tilde{Z})}{|N(\tilde{Z})|}\right) + s\tilde{R}_A(\tilde{Z})\sqrt{\tilde{R}_A(\tilde{Z})^2 - N^2(\tilde{Z})} - \tau(\tilde{Z}) \tag{41}$$

D. Propagation delay

The just-formed plasma on the initial surface is at a low temperature and is weakly ionized. The current flowing through it has to begin from nearly zero and grow to a sufficient value until the magnetic pressure on it becomes strong enough to detach it from the initial surface and begin its supersonic propagation. During this period, a significant amount of energy is spent on dissociation and ionization of the neutral gas to form a plasma. The lift-off time interval $t_L$ between the start of current and start of supersonic propagation is characterized by a negligible movement of the plasma from its initial surface. The current flow during this period is governed by a circuit equation that treats the plasma focus as a constant inductance in series with a constant resistance. When the power source is equivalent to a capacitor bank, this current has the following standard waveform

$$I_s(t) = \frac{I_0}{\sqrt{1-\lambda^2}} \exp(-\gamma_0 t) \sin(\omega_0 t); \; 0 \le t \le t_L \tag{42}$$

$$I_0 = V_0 \sqrt{\frac{C_0}{L_0}} \; ; \; \gamma_0 = \frac{R_0}{2L_0} \; ; \; \lambda \equiv \frac{R_0}{2}\sqrt{\frac{C_0}{L_0}} \; ; \; \omega_0 = \frac{\sqrt{1-\lambda^2}}{\sqrt{L_0 C_0}} \; ; \; \frac{\gamma_0}{\omega_0} = \bar{\lambda} \equiv \lambda/\sqrt{1-\lambda^2} \qquad (43)$$

where $L_0$, $C_0$ and $R_0$ are the constant inductance, capacitance and resistance of the circuit.

The scaling velocity at the initial surface has a very small, nonzero initial value at a current just sufficient to maintain the plasma ionization. It grows in proportion to the current according to equation (42). When the plasma is hot enough and has sufficient magnetic pressure pushing it, it takes off in a supersonic propagation [43].

The problem of supersonic propagation of the plasma driven by the energy being deposited by the current has been formulated [44] in terms of the laws of conservation of mass, momentum and energy expressed in a coordinate system attached with the plasma under the assumption of local planarity, neglecting curvature effects in comparison with gradient effects. This is effectively the second subproblem mention in Section II(A), where the curvature effects accounted for in the first subproblem treated in Section II(B) and (C) are neglected in the first iteration. The analysis, similar to a detonation wave problem, reveals the existence of a lower bound $v_{LB}$ for the scaling velocity proportional to the square-root of the total specific energy $\varepsilon_{d+i}$ for dissociation and ionization ($\varepsilon_{d+i} \sim 1.7 \times 10^9$ J/kg for hydrogen), below which, supersonic propagation is not possible:

$$v_0 \geq v_{LB} = f_{LB} \sqrt{\varepsilon_{d+i}} \qquad (44)$$

The factor $f_{LB}$ in (44) needs to be empirically determined as an exact specification of how hot the plasma really needs to be to take off. Its theoretical calculation should be possible via the solution of equations governing the scaled plasma variables, the second subproblem mentioned in Section II(A). Preliminary empirical estimates [40] suggest that $f_{LB} \sim 0.15$, with $\varepsilon_{d+i} = 8.45 \times 10^8 \, (\text{J/kg})$ for deuterium in a conventional Mather type plasma focus. The same value of $f_{LB}$ can be assumed for hydrogen, with $\varepsilon_{d+i} = 2 \times 8.45 \times 10^8 \, (\text{J/kg})$.

From (1), (4), (44), one gets the following relation between the maximum radius $R_I$ of the plasma formation region and the operating gas and its mass density that can be used to determine $t_L$:

$$v_{LB} = \frac{1}{\sqrt{2\mu_0 \rho_0}} \frac{\mu_0 I_0}{2\pi R_I} \tilde{I}(t_L) \qquad (45)$$

For a given $R_I$, (45) has no solution for

$$\rho_0 > \rho_{0,max} = \frac{1}{2\mu_0}\left(\frac{1}{v_{LB}}\frac{\mu_0 I_0}{2\pi R_I}\right)^2 \tilde{I}_{max}^{\;2} \tag{46}$$

where

$$\tilde{I}_{max} \equiv \text{Max}\left[\exp(-\gamma_0 t)\sin(\omega_0 t)\right] = \exp\left(-\frac{\lambda \text{ArcCos}(\lambda)}{\sqrt{1-\lambda^2}}\right) \tag{47}$$

This represents the gas pressure above which the plasma remains attached with the insulator throughout the discharge.

The scaled current at the reduced lift-off time $\bar{t}_L \equiv \omega_0 t_L$ is given by

$$\tilde{I}(\bar{t}_L) \equiv \frac{I_s(\bar{t}_L)}{I_0} = \frac{\exp(-\bar{\lambda}\bar{t}_L)\sin(\bar{t}_L)}{\sqrt{1-\lambda^2}} \tag{48}$$

The charge flow until time $t_L$ is given by

$$Q_L = Q_0\left\{1 - \exp(-\bar{\lambda}\bar{t}_L)\left(\cos(\bar{t}_L) + \bar{\lambda}\sin(\bar{t}_L)\right)\right\} \tag{49}$$

The energy dissipated in $R_0$ in time $t_L$ is found from

$$\begin{aligned}E_{R-Lift} &= R_0\int_0^{t_L} I_s^{\;2}(t)\,dt \\ &= \tfrac{1}{2}C_0V_0^{\;2}\frac{4\lambda}{(1-\lambda^2)^{3/2}}\int_0^{\bar{t}_L}\left(\exp(-2\bar{\lambda}\bar{t})\sin^2(\bar{t})\right)d\bar{t}\end{aligned} \tag{50}$$

A similar calculation can be performed for the determination of $t_L$ for any other kind of well-characterized pulsed power system, that might be used in an advanced spacecraft.

From (4),

$$\tau = Q_m^{-1}\int_{t_L}^{t} I(t')\,dt' \tag{51}$$

E. Circuit equation and model closure.

The power source is assumed to be equivalent to a capacitor bank in the subsequent discussion, although similar calculations are possible with any other well-characterized pulsed power source. The circuit schematic for the GPF problem is shown in Fig 1.

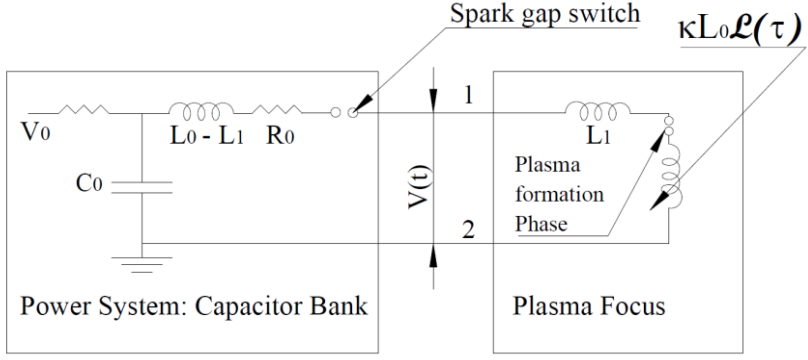

Fig. 1. Schematic of the circuit. The power system is shown as an LCR circuit with constant parameters $C_0$, $R_0$ and a part of the total static inductance of the circuit $L_0 - L_1$. It could be replaced with any other kind of pulse power source such as an inductive storage. The plasma focus is shown as an inductive element with static inductance $L_1$ and a time-varying inductance $\kappa L_0 L(\tau)$. The static inductance $L_1$ is included in the total static inductance $L_0$ of the circuit. The plasma formation phase is represented as a small spark gap switch in series with the inductance.

Define the following dimensionless quantities and scaling factors:

$$\varepsilon \equiv Q_m / C_0 V_0 \,;\, \kappa \equiv \mu_0 a / 2\pi L_0 \,;\, I_0 \equiv V_0 \sqrt{C_0 / L_0} \,;\, \tilde{I}(\tau) \equiv I(\tau(t)) / I_0 \,;\, \gamma \equiv R_0 \sqrt{C_0 / L_0} = 2\lambda \,;$$

$$\alpha = 1 - Q_L / Q_0 = \exp(-\bar{\lambda} \bar{t}_L)\left(\cos(\bar{t}_L) + \bar{\lambda} \sin(\bar{t}_L)\right); \quad T_{1/4} = \frac{\pi}{2\omega_0} \tag{52}$$

It can be shown [40] that the voltage across the plasma focus device, defined as the region over which the GV surface propagates starting from the initial surface, is given by

$$V(t) = \frac{\kappa}{\varepsilon} V_0 \left\{ \frac{d}{d\tau}\left(\frac{1}{2}\tilde{I}^2(\tau) L(\tau)\right) + \frac{1}{2}\tilde{I}^2(\tau)\frac{d}{d\tau} L(\tau) \right\} \tag{53}$$

where $L(\tau) = \int_\Omega \frac{d\tilde{r} d\tilde{z}}{\tilde{r}} \tag{54}$

and the domain of integration $\Omega$ is the region in $(\tilde{r}, \tilde{z})$ space bounded by the GV surface and the anode and cathode profiles.

The voltage across the terminals of the capacitor bank is given by standard circuit theory

$$V(t) = V_0 - C_0^{-1} \int_0^t I(t') dt' - L_0 \frac{dI(t)}{dt} - R_0 I(t) \tag{55}$$

which can be reduced using (52) to

$$V(t) = V_0 \left( \alpha - \varepsilon\tau - \varepsilon^{-1}\tilde{I}(\tau)\frac{d\tilde{I}(\tau)}{d\tau} - \gamma\tilde{I}(\tau) \right) \tag{56}$$

The voltage across the plasma focus given by (53) is in series with the voltage across the capacitor bank giving a loop voltage equal to zero. This can be rearranged as

$$\tilde{I}(\tau)\frac{d}{d\tau}\{(1+\kappa L(\tau))\tilde{I}(\tau)\} = \varepsilon(\alpha - \varepsilon\tau - \gamma\tilde{I}(\tau)) \tag{57}$$

Multiplying both sides by $(1+\kappa L(\tau))$ and introducing $\Phi(\tau) \equiv (1+\kappa L(\tau))\tilde{I}(\tau)$ this becomes

$$\frac{d\Phi^2(\tau)}{d\tau} = 2\varepsilon\alpha - 2\varepsilon^2\tau + 2(1-\alpha)\varepsilon\kappa L(\tau) - 2\varepsilon^2\kappa\tau L(\tau) - 2\varepsilon\gamma\Phi(\tau) \tag{58}$$

Equation (58) can be solved by the method of successive approximations by treating $\Phi$ as the limit of a sequence of functions $\{\Phi_0, \Phi_1 \cdots \Phi_n \cdots\}$ for large n, which satisfy the equation

$$\Phi_n^2(\tau) - \Phi_n^2(0) = 2\varepsilon\alpha\tau - \varepsilon^2\tau^2 + 2\alpha\varepsilon\kappa M_0(\tau) - 2\varepsilon^2\kappa M_1(\tau) - 2\varepsilon\gamma\int_0^\tau \Phi_{n-1}(\tau')d\tau' \tag{59}$$

$$M_0(\tau) \equiv \int_0^\tau L(\tau)d\tau;\ M_1(\tau) \equiv \int_0^\tau \tau L(\tau)d\tau \tag{60}$$

where $\Phi_0(\tau)$ is given by

$$\Phi_0^2(\tau) = \Phi_0^2(0) + 2\varepsilon\alpha(\tau + \kappa M_0(\tau)) - \varepsilon^2(\tau^2 + 2\kappa M_1(\tau)) \tag{61}$$

The integration constant $\Phi_n^2(0) = \Phi_0^2(0)$ can be calculated using

$$\Phi_0^2(0) = (1+\kappa L(0))^2 \tilde{I}^2(0) = I^2(t_L)/I_0^2 = \frac{\exp(-2\bar{\lambda}\bar{t}_L)\sin^2(\bar{t}_L)}{(1-\lambda^2)} \tag{62}$$

The sequence converges in 2-3 iterations [40] because $\varepsilon\gamma$ is a small parameter.

For evaluating the inductance, the inverse of the anode profile definition is used, giving the axial coordinate on the anode profile as a function of radial coordinate, $\tilde{Z}_A(\tilde{r})$, instead of the anode radius function $\tilde{R}_A(\tilde{Z})$ given in (29). Then,

$$L(\tau) \equiv \int_{\Omega(\tau)} \frac{d\tilde{r}d\tilde{z}}{\tilde{r}} = \int_{\tilde{r}_{min}}^{R_c} d\tilde{r} \frac{(\tilde{z}_{GV}(\tilde{r},\tau) - \tilde{Z}_A(\tilde{r}))}{\tilde{r}} \tag{63}$$

From this, the flux function $\Phi_n(\tau)$ is found from (59) and the dimensionless current is found from

$$\tilde{I}_n(\tau) = \frac{\Phi_n(\tau)}{(1+\kappa L(\tau))} \tag{64}$$

Then the real time t normalized to $T_{1/4} \equiv \pi/2\omega_0$ is found from the equation

$$\tilde{t}(\tau) \equiv t/T_{1/4} = \frac{2}{\pi}\bar{t}_L + \sqrt{1-\lambda^2}\,\frac{2\varepsilon}{\pi}\int_0^\tau \frac{d\tau}{\tilde{I}(\tau)} \tag{65}$$

The complete current waveform is then

$$\begin{aligned}
I(\tilde{t}) &= \frac{I_0}{\sqrt{1-\lambda^2}}\exp(-\pi\bar{\lambda}\tilde{t}/2)\sin(\pi\tilde{t}/2) \quad \text{for } 0 < \tilde{t} < \frac{2}{\pi}\bar{t}_L \\
&= I_0 \frac{\Phi(\tau)}{(1+\kappa L(\tau))} \quad \text{for } \frac{2}{\pi}\bar{t}_L < \tilde{t}(\tau); 0 \leq \tau \leq \tau_p
\end{aligned} \tag{66}$$

This determines the current and the shape and location of the GV surface at any time, and therefore all the scaling parameters, providing model closure. This has been illustrated with a concrete example in the tutorial paper [40].

More importantly, this analysis enables determination of the partition of stored energy. The fraction of energy in the capacitor bank at any time after the lift-off is

$$\eta_C(\tau) = (\alpha - \varepsilon\tau)^2 \tag{67}$$

The fraction of energy stored in the magnetic field is

$$\eta_M(\tau) = \frac{\Phi^2(\tau)}{(1+\kappa L(\tau))} \tag{68}$$

The fraction of energy dissipated in the resistance is

$$\eta_R = \left(\frac{1}{2}C_0V_0^2\right)^{-1}\int_0^t R_0 I^2 dt = 2\gamma\varepsilon\int_0^\tau \tilde{I}d\tau + \frac{4\lambda}{(1-\lambda^2)^{3/2}}\int_0^{\tilde{t}_L}\left(\exp(-2\bar{\lambda}\tilde{t})\sin^2(\tilde{t})\right)d\tilde{t} \tag{69}$$

The remaining fraction resides in the kinetic energy of the plasma

$$\eta_K = 1 - (\eta_C + \eta_M + \eta_R) \tag{70}$$

These relations can be used to define conditions for an optimum energy transfer from the pulse power source to the moving plasma at a predefined time $\tau_{pd}$ which could be the time $\tau_p$ to reach the axis or something similar. At this time, the capacitor storage must be fully discharged. This condition can be expressed in terms of (67) as

$$\varepsilon_{opt} = \alpha \tau_{pd}^{-1} \tag{71}$$

One could also desire the fraction of energy present in the magnetic field to be $\eta_{m,des}$ at the same time. Then, (61), (68) and (71) give, to zeroth order in the small parameter $\varepsilon\gamma$,

$$\kappa_{opt} = \frac{\eta_{m,des} - \Phi_0^2(0) - \alpha^2}{\left(2\alpha^2 \tau_{pd}^{-1} M_0(\tau_{pd}) - 2\alpha^2 \tau_{pd}^{-2} M_1(\tau_{pd}) - \eta_{m,des} L(\tau_{pd})\right)} \tag{72}$$

This demonstrates the existence of an energy transfer optimization criterion for the Generalized Plasma Focus problem. A particular configuration of the GPF such as the Mather type plasma focus, defined by its scaled geometry, thus has optimum values of the dimensionless parameters $\varepsilon$ and $\kappa$ and therefore an optimum value of $\tilde{I}(\tau_p)$.

### III. Comparison of the GPF predictions with experiments:

Even though the development of the theory of the Generalized Plasma Focus problem described above is seen to proceed logically from first principles, it is legitimate to inquire how its predictions compare with known experimental facts. Of particular importance is the following set of observations concerning near constancy of certain parameters, deduced [45-49] from a wide range of experiments [50-61] spanning the energy range from 0.1 J to 1 MJ:

(a) The pinch radius $r_p \sim (0.1-0.2)a$, the pinch height $z_p \sim (0.8-1.0)a$

(b) The mean axial velocity $\langle v_{ax} \rangle \sim 5\times 10^4 \, m/s$, the final axial velocity $\langle v_{ax,f} \rangle \sim 1\times 10^5 \, m/s$

(c) The mean radial velocity $\langle v_r \rangle \sim 1\times 10^5 \, m/s$, final radial velocity $\langle v_{r,f} \rangle \sim 2\times 10^5 \, m/s$

(d) The cross-section averaged electron density of the pinch $\langle n \rangle \sim 18 n_0$

(e) The energy density parameter $28E/a^3 \sim 5\times 10^{10} \, J/m^3$

(f) The driver parameter $I/a\sqrt{p} \sim 77 \, kA/cm-mbar^{1/2}$

(g) The magnetic field at the pinch radius $\sim 30\text{-}40 \, T$

The existence of scaling laws that apply to a whole class of plasma focus devices is a reflection of the fact that the problem of solving the equations governing plasma dynamics can be decomposed into two weakly interdependent subproblems as discussed in Section II(A), the second of which is completely independent of all information pertaining to a device and is thus universally applicable to the whole class of devices. The important question is: why are these parameters constant? Their specific values could either be a result of fundamental physics or a consequence of certain design practices or a combination thereof.

It can be observed that many of these parameters are internally related. Thus, the magnetic field at the pinch radius is related to the energy density parameter. The radial and axial velocities and the drive parameter are related. The constancy of these parameters must be explained in terms of one or more constant reference numbers that are external to the device.

An answer to this question is described below, using some results described in more detail in a tutorial paper [40]. The theoretical predictions concerning the pinch radius, pinch height and density are discussed with reference to actual measurements in two specific experiments.

Since most of the cited experiments are done with the classical Mather type geometry, the anode and insulator profiles are simple cylinders, with anode radius equal to the scale length 'a' and normalized anode height $\tilde{z}_A$. For this geometry, $\tilde{r}_{last} = 1$ referring to the shoulder of the anode profile, $\tau_{last}$ is the rundown time $\tau_R$ and $\tau_p = \tau_R + 1$. The discussion below continues to use the symbols $\tau_{last}$ and $\tilde{r}_{last}$ in the interest of generality.

(a) The pinch phase

Recall that the scaled plasma parameters such as density, velocity, pressure and magnetic field are expected to be functions of $\psi(\tau,\tilde{r},\tilde{z})$. Pending the development of a theory of spatial structure, one could model the front and rear boundaries [62] of the density distribution, representing the shock wave (SW) and the magnetic piston (MP) respectively, by two surfaces $\psi(\chi\tau,\tilde{r},\tilde{z}) = 0$ and $\psi(\tau,\tilde{r},\tilde{z}) = 0$ where $\chi$ is a constant ad hoc parameter slightly greater than unity, which makes the front boundary (shock wave) move slightly faster than the rear boundary (magnetic piston) as required by the conservation laws. The radius of the MP and SW surfaces at the anode surface at time $\tau$ are given by (39)

$$\tilde{r}_{MP}^2 = \left|\tilde{r}_{last}^2 + s(\tau - \tau_{last})\right|; \tilde{r}_{SW}^2 = \left|\tilde{r}_{last}^2 + s\chi(\tau - \tau_{last})\right|; \tag{73}$$

The MP and SW reach the axis at times

$$\tau_{p,MP} = \tau_{last} + \tilde{r}_{last}^{\,2}; \quad \tau_{p,SW} = \tau_{last} + \chi^{-1}\tilde{r}_{last}^{\,2}; \quad \tau_{p,SW} = \tau_{p,MP} + \tilde{r}_{last}^{\,2}\left(\chi^{-1} - 1\right) \tag{74}$$

For time $\tau_{p,MP} > \tau > \tau_{p,SW}$, the shock wave gets reflected from the axis forming AEF and REF branches where the radius of the REF branch at the flat top of the anode is given by (73) as

$$\tilde{r}_{SW}^{\,2} = \chi(\tau - \tau_{last}) - \tilde{r}_{last}^{\,2} \tag{75}$$

In this period the radius of the radially imploding MP at the anode surface is given by

$$\tilde{r}_{MP}^{\,2} = \tilde{r}_{last}^{\,2} - (\tau - \tau_{last}) \tag{76}$$

The REF of the shock wave and inward moving magnetic piston intersect at the anode surface at

$$\tau_{int} = \tau_{p,MP} - \vartheta, \quad \tilde{r}_{int} = \sqrt{\vartheta} \qquad \vartheta \equiv \tilde{r}_{last}^{\,2} \cdot (\chi - 1)/(\chi + 1) \tag{77}$$

From this instant onwards, they continue intersecting at distances further from the anode surface (see Fig 9 of Ref [40]) until at $\tau_{p,MP} = \tau_{last} + \tilde{r}_{last}^{\,2}$ the MP reaches the centre of the anode and itself starts getting reflected forming an REF whose radius at the anode surface is given by

$$\tilde{r}_{MP}^{\,2} = (\tau - \tau_{last}) - \tilde{r}_{last}^{\,2} = \tau - \tau_{p,MP} \tag{78}$$

which reaches $\tilde{r}_{int}$ at

$$\tau_E = \tau_{p,MP} + \vartheta \tag{79}$$

From $\tau_{p,MP}$ to $\tau_E$, the outward-moving REF of the shock wave front intersects with the inward moving CN portion of the magnetic piston front, adding the point of intersection to the zero-velocity boundary of the pinch.

At the time $\tau_E$, the axial position of the AEF of the SW is given by the relation (40),

$$\begin{aligned}\tilde{z}_{APF,SW}(\tau_E) &= 1.70246(\chi\tau_E - \chi\tau_{p,SW})^{0.443703} \\ &= 1.70246\left(\chi(\tau_{p,MP} + \vartheta) - \chi\left(\tau_{p,MP} + \tilde{r}_{last}^{\,2}(\chi^{-1} - 1)\right)\right)^{0.443703} \\ &= 1.70246\left(\chi\vartheta + \tilde{r}_{last}^{\,2}(\chi - 1)\right)^{0.443703}\end{aligned} \tag{80}$$

This is the apparent maximum normalized height of the stationary pinch before it begins its re-expansion.

The set of points of intersection between the reflected shock wave surface and the imploding magnetic piston surface define the zero-velocity "equilibrium" boundary of the plasma according the slug model of Potter [63] which is also used in the Lee model [35]. Fig 2 describes the evolution of this boundary, which is described in more detail in the tutorial paper [40].

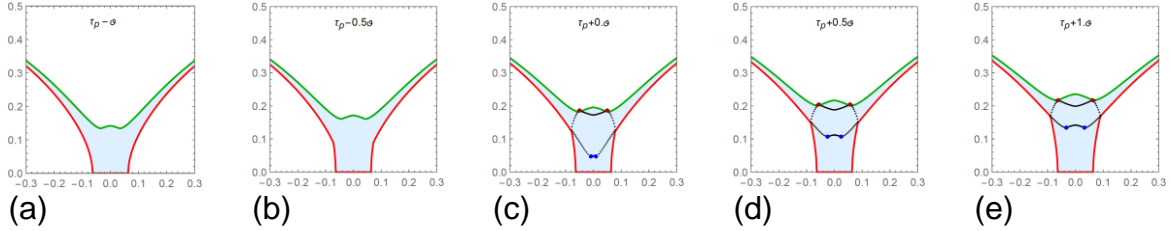

Fig 2: Evolution of equilibrium pinch boundary. (a) $\tau_{p,MP} - \vartheta$ (b) $\tau_{p,MP} - 0.5\vartheta$ (c) $\tau_{p,MP}$ (d) $\tau_{p,MP} + 0.5\vartheta$ (e) $\tau_{p,MP} + \vartheta$. For this figure, $\chi = 1.01$. Reproduced from Ref [40] (see for details) under a Creative Commons Attribution (CC BY) license.

Similar shapes are found in interferometric images from PF-1000 [64] and POSEIDON [65]. The calculated boundary at $\tau_E$ is of special significance because at this time, the REF of the MP reaches the zero-velocity boundary at the $\tilde{r}_{int}$ after reflection from the centre to restart the expansion of the pinch. This configuration, shown in Fig 2 (e), has the least radius at the anode surface with the largest height and can be compared with Fig 2(a) (at 3 ns) of Ref 64 where the radius and height of the plasma column are 8 mm and 30 mm respectively. Its compact size as compared with its preceding (-27 ns) and succeeding (33 ns) images, its temporal closeness (3 ns) to the current derivative minimum and the noticeable axial taper suggest that it corresponds to the time $\tau_E$. With the anode radius of 115 mm, the observed normalized radius and height of the pinch are 0.07 and 0.26 respectively. Equating the normalized height calculated from the model and given by (80) with the observed normalized pinch height yields $\chi = 1.00964$. For this value of $\chi$, formula (77) gives $\tilde{r}_{int} \approx 0.0695$, in surprisingly good agreement with the experimental normalized radius of 0.07.

Another example is Fig 2 of Ref 65, which depicts Abel inversion of the pinch phase of POSEIDON for shot #6308. The radius of the plasma stem is ~8 mm and the height is ~28 mm. The anode radius is 65.5 mm. The normalized plasma height is ~0.427, while the normalized plasma radius is ~0.122. For this value of normalized height, (80) gives $\chi = 1.02938$ for which (77) gives $\tilde{r}_{int} \sim 0.120328$, again a very good agreement. Note that the scaling model proposed by Lee and Serban [45] provides the values of scaled radius and height as 0.12 and 0.8 for all plasma focus devices.

The SW and MP surfaces can be considered to be locally planar [66] and equivalent to a planar, strong, piston-driven shock for which the following relations from the theory of strong shocks can be borrowed [63], with the adiabatic index $\gamma_{ad}$

$$v_{shock}/u_{piston} = \tfrac{1}{2}(\gamma_{ad}+1); \frac{\rho_{shock}}{\rho_0} = \left(\frac{\gamma_{ad}+1}{\gamma_{ad}-1}\right) \tag{81}$$

Since $v_{shock}/u_{piston} = \chi$, an effective adiabatic index can be defined from (81) as $\gamma_{eff} = 2\chi - 1$ from which the density ratio can be estimated. For the PF-1000 example [64], $\gamma_{eff} = 1.01927$, giving $\rho_{shock} \approx 105\rho_0$. At 100 Pa filling pressure at 20°C [64], the atomic number density of D$_2$ gas is $\sim 4.94 \times 10^{16}$ cm$^{-3}$ so the expected electron number density from (81) is $n_e \sim 5.2 \times 10^{18}$ cm$^{-3}$ which agrees with the peak electron density in the pinch at 3 ns in Fig 2(a) of [64] within 15%. For POSEIDON, $\gamma_{eff} = 1.05877$, giving $\rho_{shock} \approx 35\rho_0$. The filling pressure of 5 hPa of D$_2$ at 20°C corresponds to an atomic number density of $\sim 2.47 \times 10^{17}$ cm$^{-3}$. This yields for POSEIDON, the electron number density $n_e \sim 8.64 \times 10^{18}$ cm$^{-3}$ which agrees with the peak electron number density of Fig 2 of [65] within 15%. The empirical scaling $\langle n \rangle \sim 18 n_0$ mentioned above is seen to be not as good as the theoretical predictions derived above in these two examples.

Such high compression ratios and values of effective adiabatic coefficient so close to unity are features of a sequence of shocks of increasing strength [67] which is used to approximate an accelerating piston driving a non-steady shock. The term $\partial_\tau \mathrm{Log}(\tilde{I}(\tau))$ in the dimensionless equation of motion (5) and other equations governing scaled plasma variables represents such an accelerating piston and probably governs the value of $\chi$.

Using the above results, it is possible to calculate a synthetic interferogram [40]. Assuming that the laser beam travels along the x-direction, the phase shift recorded on the image plane parallel to the (y, z) plane would be

$$\varphi(y,z) = \frac{\pi a}{\lambda n_c} \int_{-1}^{1} n(\tilde{x},\tilde{y},\tilde{z}) d\tilde{x} = 2.82 \times 10^{-21} \cdot a(m) \lambda(\mu m) \int_{-1}^{1} n(\tilde{x},\tilde{y},\tilde{z}) d\tilde{x} \tag{82}$$

where $n_c$ is the critical plasma density for the laser wavelength. The theoretical density profile $n_t(\tilde{r},\tilde{z})$ can be described as

$$\frac{\rho_{shock}}{m_i} \equiv 4.8286 \times 10^{20} \, (m^{-3}) \cdot p_{D_2}(Pa) \left( \frac{\gamma_{eff}+1}{\gamma_{eff}-1} \right) \tag{83}$$

multiplied by 1 when the point $(\tilde{r},\tilde{z})$ is within the region bounded by the SW and the MP and zero otherwise. Thus

$$\varphi(\tilde{y},\tilde{z}) = 1.3616652 \cdot p_{D_2}(Pa) \left( \frac{\gamma_k+1}{\gamma_k-1} \right) \cdot a(m) \lambda(\mu m) \int_{-1}^{1} \mathbf{B}(\tilde{x},\tilde{y},\tilde{z}) d\tilde{x}$$

$$\mathbf{B}(\tilde{x},\tilde{y},\tilde{z}) = 1 \quad \text{if } \tilde{r}_{SW}(N,\tau) \leq \sqrt{\tilde{x}^2+\tilde{y}^2} \leq \tilde{r}_{mp}(N,\tau) \quad \text{and} \quad \tilde{z}_{SW}(N,\tau) \geq \tilde{z} \geq \tilde{z}_{mp}(N,\tau) \tag{84}$$

$$= 0 \quad \text{otherwise}$$

For the PF-1000 example cited above [64], the fill pressure is 100 Pa, $\lambda(\mu m) = 0.527$, $a(m) = 0.115$. The following figure 3 is a numerically constructed image of $\sin^2 \varphi(\tilde{y},\tilde{z})$.

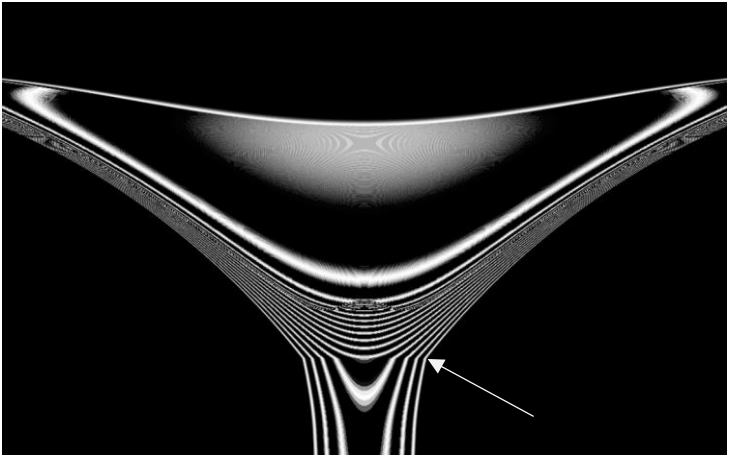

Fig 3: Numerically constructed image of $\sin^2 \varphi(\tilde{y},\tilde{z})$ using Mathematica® Image function. The noticeable change in the slope of the fringe at the place marked by the arrow is also seen in the experimental interferograms [64, 65]. Reproduced from Ref [40] (see for details) under a Creative Commons Attribution (CC BY) license.

The above examples suggest that the observation that the normalized pinch radius and pinch height are nearly constant for all plasma focus devices is an aspect of the device-independent dynamics of scaled plasma variables, the subproblem 2 mentioned in Section II(A), which also apparently governs the ratio of the pinch density to the fill density.

(b) Universality of scaling

The scaling parameter for velocity is simply proportional to the drive parameter as originally defined by Lee and Serban [45],

$$v_0 \equiv \frac{\sqrt{\mu_0}}{2\sqrt{2}\pi} \frac{\tilde{I}(t)}{\tilde{r}} \left( \frac{I_0}{a\sqrt{\rho_0}} \right) \quad (85)$$

The average kinetic energy per particle of mass $m_i$ has a scale of

$$E_k = \frac{1}{2} m_i v_0^2 \quad (86)$$

This must be large enough to create a fully dissociated and ionized plasma. Considering that for hydrogen the dissociation energy is 4.48 eV and ionization energy is 13.6 eV, the kinetic energy (86) should be of the order of 18 eV. For deuterium, the corresponding velocity comes out as ~4.2 x $10^4$ m/sec, in good agreement with the average axial velocity estimate. This is related to the velocity below which, propagation of the sheath is not allowed by conservation laws [44].

The velocity $1 \times 10^5$ m/s corresponds to deuterium kinetic energy ~100 eV, which is the order of electron temperature observed by XUV spectroscopy [68]. From this scale, energy concentrating mechanisms operating via instabilities can reach ion kinetic energies ~ 1 keV where appreciable fusion reaction rates become accessible. Devices optimized for neutron emission would work closer to this value.

The scaling parameter $E_0$ for the electric field $E_0 = v_0 B_0 = v_0^2 \sqrt{2\mu_0 \rho_0}$ has an upper bound [44] given by the electric breakdown strength of the gas $E_b(H_2) = 8.21 \times 10^{21} \rho_0 / m_i$. This bound provides the so-called ionization stability condition [66] that ensures that ingestion of neutral gas into the moving plasma occurs only at the interface between the shock wave and the neutral gas. Since there is an upper bound on the mass density of the fill gas for a particular device given by (46), there is an upper bound also on $v_0$

$$v_0^2 < 8.21 \times 10^{21} \left( \frac{1}{v_{LB}} \frac{I_0}{4\pi m_i R_I} \right) \tilde{I}_{max}(t_L) \quad (87)$$

The upper bound on both $v_0$ and $\rho_0$ translates into an upper bound on $B_0$. The energy density of the magnetic field is the scaling parameter for pressure, which can be expressed as

$$p_0 \equiv \frac{B_0^2}{2\mu_0} = \frac{1}{2} \frac{C_0 V_0^2}{a^3} \frac{\kappa}{2\pi} \frac{\tilde{I}^2(\tau)}{\tilde{r}^2} ; \quad (88)$$

The upper bound on $B_0$ is seen to put a bound on the energy density parameter $C_0 V_0^2 / a^3$ for optimized plasma focus devices, in view of the fact the $\kappa$ and $\tilde{I}(\tau)$ have optimum values obtained in subsection II E.

For a given plasma focus, (87) represents a failure of sheath propagation at a sufficiently high voltage [3] since the left-hand side of the inequality increases as the square of the voltage while the right hand side increases linearly as the voltage. As a given plasma focus device is operated at incrementally higher voltages, the violation of condition (87) would cause electron multiplication in the neutral gas layer because of electric field induced electron impact ionization and not by the shock-heated plasma. This would cause an ionization wave to form and move ahead of the shock wave. It would carry the current up to the axis but not the mass associated with the shock wave. There would thus be very little deuterium to cause fusion reactions using the energy transported up to the axis by the ionization wave.

This probably accounts for the failure of the neutron yield scaling in large plasma focus devices. Inequality (87) also suggests that this failure can be mitigated by decreasing the insulator radius, for example, by making the insulator radius much less than the anode radius. It also suggests that the proposed space propulsion concept, outlined in the Introduction and further discussed below, would not be affected by it, since the target density is decoupled from the density in the lift-off region.

This explains why the drive parameter varies in a limited range for plasma focus devices varying in physical size and energy storage over many orders of magnitude. In physical terms, too low a value of the drive parameter does not provide enough kinetic energy per particle to cause sufficient ionization and therefore fails to launch the plasma focus sheath. Too high a value leads to electric breakdown ahead of the sheath causing the current to flow behind a faster ionization front that detaches from the plasma and races ahead of it. The ionization front transports the electrical energy but not the mass and therefore fails to form a dense plasma target that can accept the delivered energy and produce fusion reactions. Both the limits are dependent on material properties of deuterium.

Since most devices optimised for neutron emission would attempt to operate near the upper bound of energy per particle, the scaling parameters of velocity and magnetic field and the related drive parameter and energy density parameter for all such devices would have nearly the same value.

The above discussion indicates that the predictions of the GPF theory compare well with the experimental knowledgebase.

IV. <u>Application of the scaling theory to an illustrative space propulsion concept</u>

The attractive properties of a plasma-focus-based fusion drive for an interplanetary or deep space mission [16,17] could be realized in practice only by combining the power density amplifying properties of a plasma-focus-like device with a well-studied current-driven fusion load such as a z-pinch driven dynamic hohlraum [18, 19] or magnetized liner inertial fusion (MAGLIF) [20, 21]. However, the process of transferring energy from a pulse power source to such a load needs to be studied and optimized in a laboratory experiment, in a manner that can be credibly extrapolated to an up-scaled prototype. This section looks at a configuration that could serve this purpose.

At its most basic level, a space propulsion concept is a method of ejecting mass with a high velocity. The use of a coaxial gun to accelerate a plasma to a high velocity ($\sim 10^5$ m/sec) is well known [69,70] but this happens at a very low gas density and hence a very low mass ejection rate ($\sim 1.5 \times 10^{-10}$ kg per pulse). Instead, if a high-density metal wire can be rapidly converted into a magnetically confined plasma jet, it could be evaluated as a laboratory surrogate of a propulsion concept based on transport of energy by a GPF current sheath to a current-driven fusion load. This configuration could even be evaluated as a non-nuclear electric propulsion engine for near-earth missions.

The feasibility of such a laboratory experiment should depend on the scale of mechanical power density $\Pi_m$ (watts/m$^2$): the ability to apply a high pressure rapidly over a small concentrated area. The scaling theory of Section II suggests that this should be the product of the scale factors for pressure and velocity.

$$\Pi_m \equiv p_0 v_0 = \frac{B_0^3}{2\mu_0 \sqrt{2\mu_0 \rho_0}} \tag{89}$$

The scale of the electrical power density $\Pi_e$ into the device should be the product of the scale factors for voltage and current divided by the scale of the cross-sectional area

$$\Pi_e \equiv V_0 I_0 / \pi a^2 \tag{90}$$

Comparing the two numbers and using (52)

$$\frac{\Pi_m}{\Pi_e} = \frac{\kappa}{8\varepsilon}\left(\frac{\tilde{I}(\tau)}{\tilde{r}}\right)^3 \tag{91}$$

Since the parameters $\kappa$ and $\varepsilon$ are chosen for optimum energy transfer according to (71) and (72), it is clear from (91) that amplification of power density requires an electrode configuration that maximizes $\tilde{I}(\tau)/\tilde{r}$ at the anode surface at the moment of ejection. As the GV surface decreases in radius, its inductance increases as $\text{Log}(1/\tilde{r})$ and hence current decreases as $1/\sqrt{\text{Log}(1/\tilde{r})}$. But the ratio $\tilde{I}(\tau)/\tilde{r}$ could still increase.

This does not happen in a conventional plasma focus with a wire placed at its axis [71]. In this case, the plasma gradients are perpendicular to the wire surface. The interaction between the plasma and the wire involves partial transmission and reflection of shock waves at the plasma-wire interface. The reflected shock wave delays the arrival of the current carrying plasma layer at the wire so that hardly any current flows through the wire and the interferometric observation [71] of the wire-plasma interaction shows that the wire remains intact for quite some time.

The way to a lower radius and higher $\tilde{I}(\tau)/\tilde{r}$ at the anode surface is to orient the plasma gradients parallel to the anode, by letting the plasma slide along the anode rather than collide head-on with it as in the case of the wire placed on the axis of a conventional plasma focus.

In the proposed illustrative concept, the energy supplied by the pulse power source is transported in the form of current flowing behind the sheath of a modified plasma focus, where the anode gradually decreases in radius along its length. Its extreme end has an orifice through which a metallic wire of radius $r_{wire}(m)$, height $h_{wire}(m)$ and density $\rho_{wire}(kg/m^3)$ is continuously inserted. The GV surface, mimicking the current-carrying plasma front, slides over this anode. It ultimately flows over the wire, providing a current path that passes from the anode, through the wire, over its junction with the GV surface, to the cathode. The plasma sliding over the anode to the fuel wire essentially acts as a plasma flow switch [72]. In the event of the failure of the ionization stability condition, the role of the plasma flow switch is taken over by the ionization wave which detaches from the plasma and races ahead carrying current behind it, so that the transfer of current to the wire would not be affected.

The extreme level of current density flowing through the wire material and its high rate of rise cause phase transitions [73] that lead to formation of a superheated vapour-liquid phase mixture and ultimately also to a plasma [18]. This plasma is subject to an intense compressive magnetic pressure that causes it to be ejected axially in the form of a collimated jet with a velocity of the order of the Alfven velocity of the plasma

$$v_{jet} \sim \frac{1}{\sqrt{2\mu_0 \rho_{wire}}} \frac{\mu_0 I_0}{2\pi r_{wire}} \tilde{I}(\tau_2) \quad (92)$$

The total impulse $\mathcal{J}$ created by the axial ejection of this mass in one shot would be

$$\begin{aligned}
\mathcal{J} &\sim m_{wire} v_j \sim \rho_{wire} \cdot \pi r_{wire}^2 \cdot h_{wire} \cdot \frac{1}{\sqrt{2\mu_0 \rho_{wire}}} \frac{\mu_0 I_0}{2\pi r_{wire}} \tilde{I}(\tau_2) \\
&\sim \sqrt{\rho_{wire}} \cdot r_{wire} \cdot h_{wire} \cdot \frac{\sqrt{\mu_0}}{2\sqrt{2}} I_0 \tilde{I}(\tau_2)
\end{aligned} \quad (93)$$

The thrust would then depend on the repetition rate of the device. The parameter that governs the scale of the impulse is seen to be proportional to the current at the moment of ejection and the height $h_{wire}(m)$ that is vaporised. The latter could provide a control mechanism for the thrust, similar to fuel injection rate in an internal combustion engine.

This section looks at the application of the scaling theory developed above to the estimation of the jet velocity and impulse per shot according to relations (92) and (93) with reference to both a general scheme and a concrete numerical example relevant for a laboratory test. For the latter, a 34-gauge stainless steel hypodermic needle tube (outer diameter 0.16 mm, inner diameter 0.05 mm, density 7.8 gm/cm$^3$, length ~ 1 cm) is assumed [74]. The mass of wire ejected per shot would then be $\rho_{wire} \cdot \pi r_{wire}^2 \cdot h_{wire} \sim 6 \times 10^{-6}$ kg. The scale length a of the laboratory experiment is chosen to be 16 mm, 100 times larger than the wire diameter. The charging voltage is taken to be 20 kV, typical of a small laboratory facility. The operating gas is assumed to be hydrogen, which is easy to store and transport as a metal hydride.

The GPF problem of the plasma flow switch is illustrated for a simple tapered insulator and anode as described in the following profile in scaled geometry, illustrated in Fig 4:

$$\begin{aligned}
\tilde{R}_I(\tilde{Z}) &= \tilde{R}_I^{(0)} + \left(\tilde{R}_I^{(1)} - \tilde{R}_I^{(0)}\right) \tilde{Z}/\tilde{Z}_I^{(1)} & 0 \leq \tilde{Z} \leq \tilde{Z}_I^{(1)} &\Leftrightarrow I_1 \\
&= \tilde{R}_I^{(1)} + \left(\tilde{Z} - \tilde{Z}_I^{(1)}\right)\left(\tilde{R}_I^{(2)} - \tilde{R}_I^{(1)}\right)/\left(\tilde{Z}_I^{(2)} - \tilde{Z}_I^{(1)}\right) & \tilde{Z}_I^{(1)} \leq \tilde{Z} \leq \tilde{Z}_I^{(2)} &\Leftrightarrow I_2
\end{aligned} \quad (94)$$

$$\begin{aligned}
\tilde{R}_A(\tilde{Z}) &= \tilde{R}_A^{(1)} = \tilde{R}_I^{(2)} & \tilde{Z}_A^{(0)} \leq \tilde{Z} \leq \tilde{Z}_A^{(1)}; \tilde{Z}_A^{(0)} = \tilde{Z}_I^{(2)} = \tilde{z}_I & \Leftrightarrow \text{Stem} \\
&= \tilde{R}_A^{(1)} + \left(\tilde{Z} - \tilde{Z}_A^{(1)}\right)\left(\tilde{R}_A^{(2)} - \tilde{R}_A^{(1)}\right) / \left(\tilde{Z}_A^{(2)} - \tilde{Z}_A^{(1)}\right) & \tilde{Z}_A^{(1)} \leq \tilde{Z} \leq \tilde{Z}_A^{(2)} & \Leftrightarrow \text{Taper} \\
&= \tilde{R}_A^{(2)} & \tilde{Z}_A^{(2)} \leq \tilde{Z} \leq \tilde{Z}_A^{(3)} & \Leftrightarrow \text{Fuel wire} \\
&= \tilde{R}_A^{(2)}\left(\tilde{Z}_A^{(4)} - \tilde{Z}\right) / \left(\tilde{Z}_A^{(4)} - \tilde{Z}_A^{(3)}\right) & \tilde{Z}_A^{(3)} \leq \tilde{Z} \leq \tilde{Z}_A^{(4)} & \Leftrightarrow \text{End Cap}
\end{aligned} \quad (95)$$

The inverse function $\tilde{Z}_A(\tilde{r})$ in (63) is then

$$\begin{aligned}
\tilde{Z}_A(\tilde{r}) &= \tilde{Z}_A^{(4)} - \tilde{r}\left(\tilde{Z}_A^{(4)} - \tilde{Z}_A^{(3)}\right) / \tilde{R}_A^{(2)} & 0 < \tilde{r} \leq \tilde{R}_A^{(2)} \\
&= \tilde{Z}_A^{(1)} + \left(\tilde{r} - \tilde{R}_A^{(1)}\right)\left(\tilde{Z}_A^{(2)} - \tilde{Z}_A^{(1)}\right) / \left(\tilde{R}_A^{(2)} - \tilde{R}_A^{(1)}\right) & \tilde{R}_A^{(2)} < \tilde{r} \leq \tilde{R}_A^{(1)} \\
&= 0 & \tilde{R}_A^{(1)} < \tilde{r} < \tilde{r}_C
\end{aligned} \quad (96)$$

This method of specifying the anode and insulator profiles can be generalized to any number of segments for performance optimization.

From (36)

$$\tau_1 = 2\left(\tilde{Z}_A^{(1)} - \tilde{z}_I\right) \quad (97)$$

The current reaches the beginning of the fuel wire at

$$\tau_2 = \tau_1 + \sqrt{1 + \left(\left(\tilde{R}_A^{(2)} - \tilde{R}_A^{(1)}\right) / \left(\tilde{Z}_A^{(2)} - \tilde{Z}_A^{(1)}\right)\right)^2} \left(\tilde{Z}_A^{(2)} - \tilde{Z}_A^{(1)}\right)\left(\tilde{R}_A^{(1)} + \tilde{R}_A^{(2)}\right) \quad (98)$$

Denoting the normalized height of the taper $\tilde{h}_T \equiv \tilde{Z}_A^{(2)} - \tilde{Z}_A^{(1)}$, the normalized height of the stem $\tilde{h}_S \equiv \tilde{Z}_A^{(1)} - \tilde{z}_I$ and recognizing that the stem radius $\tilde{R}_A^{(1)} = 1$ by definition and $\tilde{R}_A^{(2)} \square \ \tilde{R}_A^{(1)}$, (98) can be approximated as

$$\tau_2 \approx 2\tilde{h}_S + \tilde{h}_T \quad (99)$$

From (35), the current moves along the surface of the fuel according to the relation

$$\tau(\tilde{Z}) = \tau_2 + 2\tilde{R}_A^{(2)}\left(\tilde{Z} - \tilde{Z}_A^{(2)}\right) \quad (100)$$

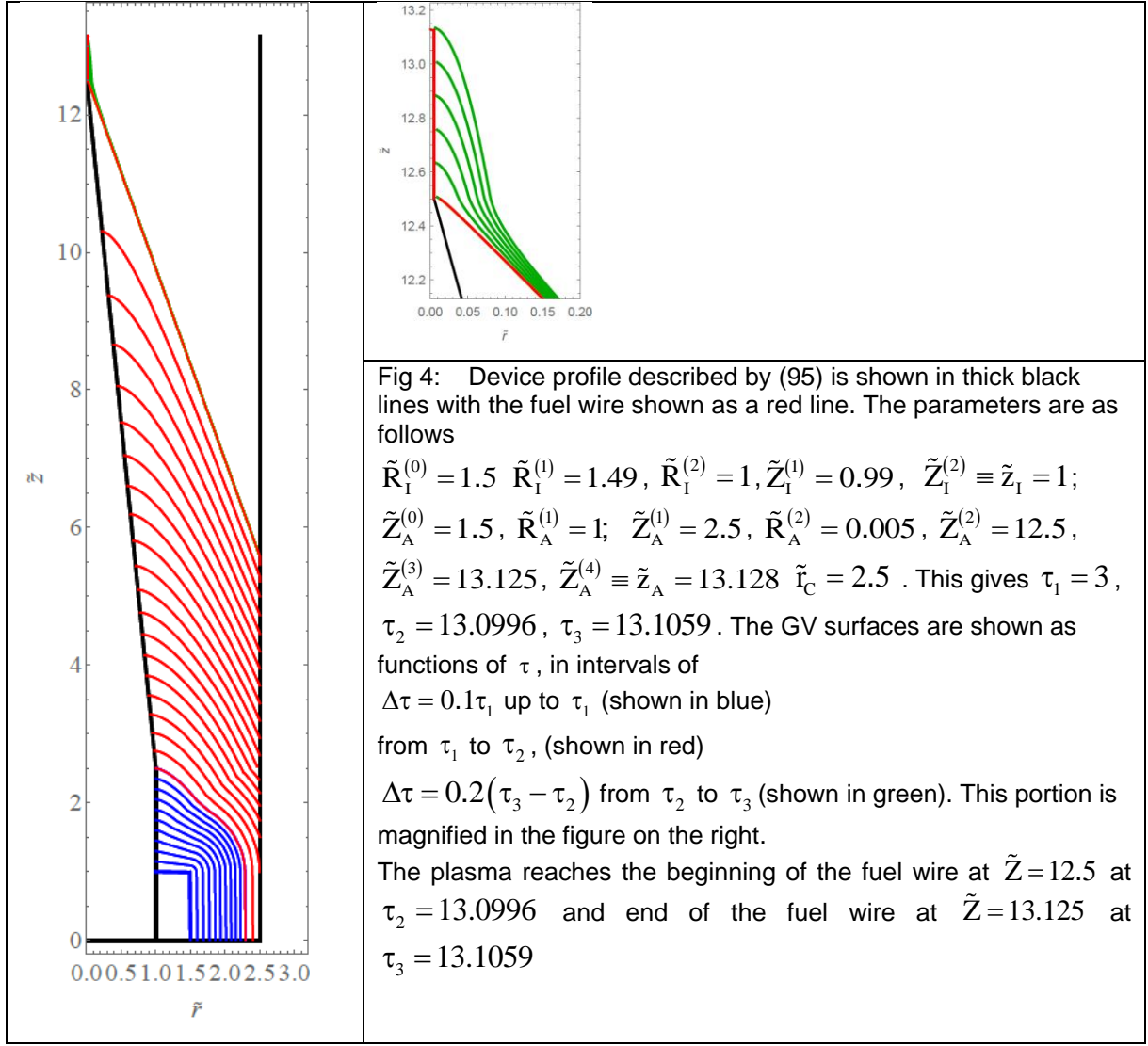

Fig 4: Device profile described by (95) is shown in thick black lines with the fuel wire shown as a red line. The parameters are as follows

$\tilde{R}_I^{(0)} = 1.5$ $\tilde{R}_I^{(1)} = 1.49$, $\tilde{R}_I^{(2)} = 1$, $\tilde{Z}_I^{(1)} = 0.99$, $\tilde{Z}_I^{(2)} \equiv \tilde{z}_I = 1$;

$\tilde{Z}_A^{(0)} = 1.5$, $\tilde{R}_A^{(1)} = 1$; $\tilde{Z}_A^{(1)} = 2.5$, $\tilde{R}_A^{(2)} = 0.005$, $\tilde{Z}_A^{(2)} = 12.5$,

$\tilde{Z}_A^{(3)} = 13.125$, $\tilde{Z}_A^{(4)} \equiv \tilde{z}_A = 13.128$ $\tilde{r}_C = 2.5$. This gives $\tau_1 = 3$, $\tau_2 = 13.0996$, $\tau_3 = 13.1059$. The GV surfaces are shown as functions of $\tau$, in intervals of

$\Delta\tau = 0.1\tau_1$ up to $\tau_1$ (shown in blue)

from $\tau_1$ to $\tau_2$, (shown in red)

$\Delta\tau = 0.2(\tau_3 - \tau_2)$ from $\tau_2$ to $\tau_3$ (shown in green). This portion is magnified in the figure on the right.

The plasma reaches the beginning of the fuel wire at $\tilde{Z} = 12.5$ at $\tau_2 = 13.0996$ and end of the fuel wire at $\tilde{Z} = 13.125$ at $\tau_3 = 13.1059$

The dimensionless dynamic inductance $\mathsf{L}(\tau)$ calculated from (63) is shown in Fig 5

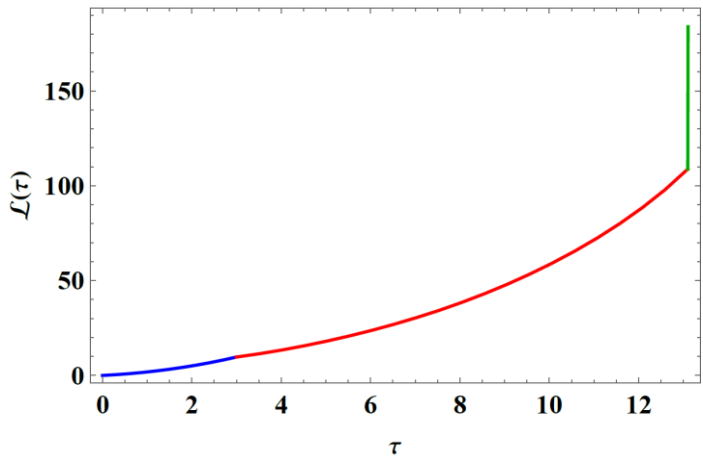

Fig 5: The variation of the dimensionless dynamic inductance $\mathsf{L}(\tau)$ for the profile of Fig. 2. The colours correspond to the colours in Fig 2. The following parameters are also calculated

$\tau_2^{-1}\mathsf{M}_0(\tau_2) = 35.6293$,

$\tau_2^{-2}\mathsf{M}_1(\tau_2) = 26.1737$,

$\mathsf{L}(\tau_2) = 108.79$.

Estimation of the impulse per shot from (93) requires calculation of the current at ejection using (66) which in turn requires values of the lift-off phase parameters $\alpha$ and $\Phi_0^2(0)$, propagation parameters $\kappa$ and $\varepsilon$ as well as the energy storage parameters $V_0, C_0, L_0, R_0$.

These are determined iteratively, first assuming the parameters of the insulator and anode profiles in (94) and (95), calculating the inductance variation and then the other parameters mentioned above. Subsequent iterations can be designed for approaching a desired outcome more closely.

The decision process that leads to these values begins with assuming a value for the parameter $\bar{\lambda}$ in (48), which is related to the ratio between successive peaks of the current /voltage waveforms (voltage reversal factor) in an unloaded capacitor bank. This factor is closely related to the life of the dielectrics involved and also to the resistive dissipation of energy in the circuit. Assuming 80% voltage reversal, $\bar{\lambda} \sim -\ln(0.8)/\pi \approx 0.0710288$, $\lambda = 0.0708503$, $\gamma = 0.141701$. Next the value of $1-\alpha$, the fraction of stored charge that can be utilized for the lift-off phase, must be chosen, (say 0.1). Relation (49) then gives $\bar{t}_L = 0.426395$ as the root of a transcendental equation. Using the values of $\lambda$ and $\bar{t}_L$, (48) gives $\tilde{I}(\bar{t}_L) = 0.426395$ and (62) gives $\Phi_0^2(0) = 0.181813$.

Next, the equivalent capacitance should be required to have zero charge by the time the current front reaches the beginning of the wire at $\tau_2$. From (67), this condition implies

$$\varepsilon = \frac{\alpha}{\tau_2} = \frac{\alpha}{2\tilde{h}_S + \tilde{h}_T} \tag{101}$$

giving $\varepsilon = 0.0687043$ for the profile shown in Fig 2.

From (68) and (101) the following expression is obtained for the value of $\kappa$ for a desired $\eta_M(\tau_2)$ using the zeroth approximation of the sequence (59).

$$\kappa = \frac{\Phi_0^2(0) + \alpha^2 - \eta_M(\tau_2)}{\left(\eta_M(\tau_2)L(\tau_2) - 2\alpha^2\tau_2^{-1}M_0(\tau_2) + 2\alpha^2\tau_2^{-2}M_1(\tau_2)\right)} \tag{102}$$

Using $\eta_M(\tau_2) = 0.7$, one gets $\kappa = 0.00479677$ for the profile in Fig 2.

From the radius 'a' of the anode, chosen to be 16 mm, the static inductance of the circuit is decided by

$$L_0 = \frac{\mu_0 a}{2\pi\kappa} \approx 667 \text{ nH} \tag{103}$$

From (45)

$$\sqrt{2\mu_0\rho_0} = \frac{1}{v_{LB}} \frac{\mu_0 I_0 \tilde{I}(\bar{t}_L)}{2\pi R_I} \tag{104}$$

The charge on the capacitor bank is therefore

$$C_0 V_0 = \frac{Q_m}{\varepsilon} = \frac{\pi \mu_0^{-1} a^2 \sqrt{2\mu_0 \rho_0}}{\varepsilon} = \frac{I_0}{v_{LB}} \frac{a^2}{2R_I} \frac{\tilde{I}(\bar{t}_L)}{\varepsilon} \tag{105}$$

The capacitance is then given by

$$C_0 = \frac{2\pi}{\mu_0} \frac{a}{v_{LB}^2} \left(\tilde{R}_I^{(0)}\right)^{-2} \tilde{I}^2(\bar{t}_L) \frac{\kappa}{4\varepsilon^2} = \frac{2\pi}{\mu_0} \frac{a}{v_{LB}^2} \left(\tilde{R}_I^{(0)}\right)^{-2} \tilde{I}^2(\bar{t}_L) \frac{\kappa(2\tilde{h}_S + \tilde{h}_T)^2}{4\alpha^2} \tag{106}$$

This relation could also be used to iteratively design a tapered profile that provides the best match to an existing capacitor bank.

Then

$$I_0 = V_0 \sqrt{\frac{C_0}{L_0}} = V_0 \frac{2\pi}{\mu_0} \frac{\kappa}{2\varepsilon} \frac{\tilde{I}(\bar{t}_L)}{v_{LB} \tilde{R}_I^{(0)}} = V_0 \frac{2\pi}{\mu_0} \frac{\kappa \tau_2}{2\alpha} \frac{\tilde{I}(\bar{t}_L)}{v_{LB} \tilde{R}_I^{(0)}} = V_0 \frac{2\pi}{\mu_0} \frac{\kappa(2\tilde{h}_S + \tilde{h}_T)}{2\alpha} \frac{\tilde{I}(\bar{t}_L)}{v_{LB} \tilde{R}_I^{(0)}} \tag{107}$$

From (104), and (107),

$$\rho_0 (kg/m^3) = (2\mu_0)^{-1} \frac{V_0^2}{4a^2} \frac{\kappa^2 (2\tilde{h}_S + \tilde{h}_T)^2}{\alpha^2} \frac{\tilde{I}^4(\bar{t}_L)}{\left(v_{LB} \tilde{R}_I^{(0)}\right)^4} \tag{108}$$

The following numbers are obtained for the assumed parameters:

$C_0 \cong 43$ μF, $I_0 \cong 160$ kA, $E_0 = \frac{1}{2} C_0 V_0^2 \cong 8.6$ kJ, $T_{1/4} \cong 8.45$ μs  (109)

The fill gas density $\rho_0 \cong 0.00342$ (kg/m$^3$) corresponds to a hydrogen pressure of ~43 millibar. The question of initial plasma formation for this pressure is discussed in Section IV.

The current waveform is shown in Fig 6

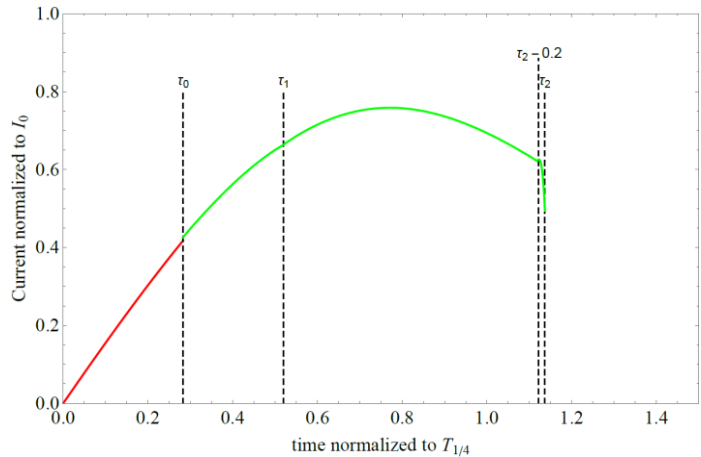

Fig 6: Current waveform according to (65) and (66) with values of α, ε and κ given by (101) and (102) . The value of γ is mentioned in the text. The red portion is the lift-off region. The green line is the current in the propagation region. The GV surface reaches the wire at $\tau_2$ given by (99). There is a sharp drop in current between $\tau_2 - 0.2$ and $\tau_2$. The normalized current $\tilde{I}(\tau)$ at $\tau_2$ is 0.50 and at $\tau_3$ it is 0.49.

However, the main feature that this plasma flow switch is meant to achieve is the ratio $\tilde{I}(\tau)/\tilde{R}_A(\tilde{Z}(\tau))$. Fig. 7 shows this in two ways: one is the ratio itself and second the resulting magnetic field at the surface of the wire.

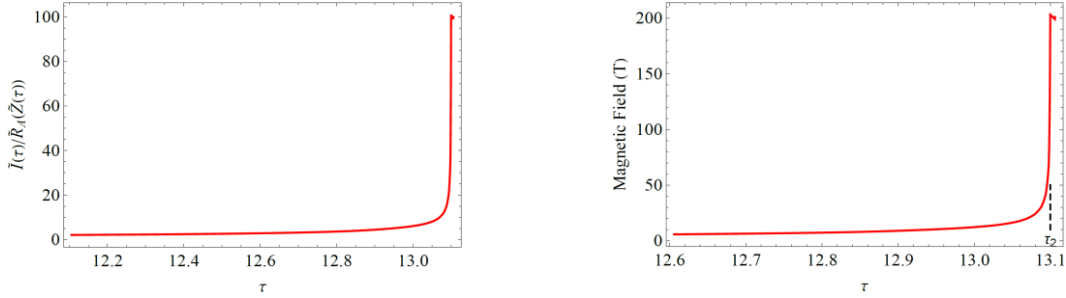

Fig 7: The ratio $\tilde{I}(\tau)/\tilde{r}$ at the anode. It peaks at $\tau_2$ to a value ~100. The power amplification according to (91) is ~9000. The magnetic field at the surface of the anode rises from 20 T to ~200T in ~ 40ns. By the time the GV surface reaches the wire at $\tau_2$, the magnetic field is already ~200 T.

The maximum power density amplification given by (91) is ~9000. The current in the wire is ~ 80 kA during the travel of the GV surface over the wire. The current density is $1.8 \times 10^{12} A/m^2$.

A metallic wire carrying current density $J(t)$ explodes at time $t_{exp}$ if the following condition is satisfied.

$$\int_0^{t_{exp}} J^2(t)dt = \bar{h} \tag{110}$$

where $\bar{h}$ is a material property called specific action integral [73] ($\bar{h} \sim 1.4 \times 10^9 A^2 \cdot s/cm^4$ for Fe). The action integral for the fuel wire becomes:

$$\text{ActionIntegral} = A^{-2}\int_{t_2}^{t_3} I^2(t)dt = A^{-2}\int_{t_2}^{t_3} I^2(t)\frac{Q_m d\tau}{I(t)} = A^{-2}Q_m I_0 \int_{\tau_2}^{\tau_3} \tilde{I}(\tau)d\tau \tag{111}$$
$$= 4.53 \times 10^{12} \cdot A^{-2} \cdot s/cm^4 \cong 3235\bar{h}$$

The wire therefore explodes in an interval $\Delta\tau \sim (\tau_3 - \tau_2)/3000$ after the arrival of the GV surface at the base of the wire and forms a plasma z-pinch [75]. Its Alfven velocity is $v_j \sim 1450 m/s$ and the impulse is $\mathcal{J} \sim 0.002$ kg-m/sec. The radial Alfven transit time over the inner radius ~ 0.025 mm is ~17 ns, while the sheath travel time $t_3 - t_2$ over the length of the wire is ~8.4 ns. The timescale of the explosion according to (110) is then $t_{exp} \sim 8.4\text{ns}/3000 \sim 3\text{ ps}$.

However, the major takeaway of this exercise is the scaling behaviour of the Alfven velocity (92), impulse (93) and the stored energy and charge:

$$v_{jet} \sim \left[\frac{\tilde{I}(\tau_2)\tilde{I}(\bar{t}_L)}{\tilde{R}_I^{(0)}\tilde{R}_A^{(2)}}\frac{\kappa(2\tilde{h}_S+\tilde{h}_T)}{2\alpha}\right]\left\{\frac{V_0}{v_{LB}}\frac{1}{a\sqrt{2\mu_0\rho_{wire}}}\right\} \tag{112}$$

$$\mathcal{J} \sim \left[\frac{\tilde{I}(\tau_2)\tilde{I}(\bar{t}_L)\tilde{R}_A^{(2)}\left(\tilde{Z}_A^{(4)}-\tilde{Z}_A^{(3)}\right)}{\tilde{R}_I^{(0)}}\frac{\kappa(2\tilde{h}_S+\tilde{h}_T)}{2\alpha}\right]\left\{\frac{V_0}{v_{LB}}\frac{\pi a^2\sqrt{\rho_{wire}}}{\sqrt{2\mu_0}}\right\} \tag{113}$$

$$E_0 = \frac{1}{2}C_0V_0^2 = \left[\left(\tilde{R}_I^{(0)}\right)^{-2}\tilde{I}^2(\bar{t}_L)\kappa\frac{(2\tilde{h}_S+\tilde{h}_T)^2}{4\alpha^2}\right]\left\{\frac{1}{2}V_0^2\frac{2\pi}{\mu_0}\frac{a}{v_{LB}^2}\right\} \tag{114}$$

$$Q_0 = C_0V_0 = \left[\left(\tilde{R}_I^{(0)}\right)^{-2}\tilde{I}^2(\bar{t}_L)\kappa\frac{(2\tilde{h}_S+\tilde{h}_T)^2}{4\alpha^2}\right]\left\{V_0\frac{2\pi}{\mu_0}\frac{a}{v_{LB}^2}\right\} \tag{115}$$

The quantities in the square brackets are dependent on the scaled geometry given by (94) and (95) while those in the braces depend on choices made on practical grounds. The impulse is seen to be proportional to the charge stored on the capacitor, rather than the energy.

Since the energy density of dielectric energy storage at a given operating stress (which is related to its operating life) is a material property, relation (114) provides a measure of the mass of the energy storage required to achieve the impulse. The fact that the operationally relevant quantity $\mathcal{I}$ depends on the stored charge rather than stored energy has implications on the choice of operating voltage, with profound downstream consequences for the choices related to pulse power technology. These relations therefore provide a toolkit for examining the thrust to weight ratio as an optimizing criterion in the parameter space.

The transport of energy by the moving plasma would also be subject to the possibility of electric breakdown of the neutral gas ahead of the sheath when $\tilde{I}(\tau)/\tilde{R}_A(\tilde{Z}(\tau))$ becomes sufficiently high. However, the fast ionization front racing ahead of the plasma sheath would simply accelerate the switching action by shortening the time required to transport the current to the fuel wire. The difference between this case and the conventional plasma focus is that the moving sheath is required to transform itself into the fusion target on reaching the axis for the latter but the proposed concept already has a fusion target situated at the axis in the form of a thin metal tube filled with fusion fuel at high density waiting for the current transfer from the moving sheath.

This exercise thus illustrates the usefulness of the GPF formalism for obtaining ballpark numbers and scaling relations relevant to a newly proposed plasma propulsion concept based on rapid explosion of a metal wire (or tube) to form an axially directed radially confined jet. The above exercise is mainly aimed at facilitating a laboratory experiment for validating the scaling theory and is not suggested as an interplanetary space propulsion concept by itself.

V. <u>Some aspects of laboratory studies of space propulsion</u>

The scaling theory resulting from the Generalized Plasma Focus problem can be applied wherever its elements are present. One of its applications is design of laboratory experiments on space propulsion whose conclusions could be credibly extrapolated to the design of a larger scale experimental study. Section IV has already demonstrated the procedure for designing such laboratory experiments, obtaining ballpark numbers which can be tweaked to suit particular requirements. This section comments on the rationale, motivation and feasibility of such laboratory studies.

The purpose of such laboratory studies would be fivefold:

a. Validation of the GPF predictions concerning the propagation of the plasma in a modified plasma focus with a tapered anode profile is necessary to provide confidence in engineering design of larger and more expensive facilities. One could, for example, measure the voltage across and current through the plasma, which could be used to calculate the inductance variation. This could be done for different profiles and profile parameters, comparing predictions and results and provide a baseline validation reference.

b. Obtaining measured numbers for the momentum and velocity of the jet and verifying their absolute magnitude and scaling is necessary in view of the advances in the science of exploding conductors [73]. The energy deposition processes at very high current density are a topic of active research in the context of wire array z-pinches.

c. The proposed concept can also be looked upon as an alternative to a wire array z-pinch for creating a dynamic hohlraum target that has a more symmetric initial configuration. In this case, the tube is filled with a high atomic number foam with an embedded ICF capsule. The experimentally determined energy deposition rate in the wall material would then be of direct relevance for determining the driver conditions.

d.  Inventing ways to form a gaseous environment in the path of the plasma flow switch that is open to an ultrahigh vacuum source – space – is going to be a crucial engineering task. Perhaps this could be done using a porous metal shell for the anode and letting gas diffuse through the surface. Non-uniform gas distributions would perhaps affect the plasma flow and affect the shape of the GV surface leading to corrections to the inductance. This would be an extension to the present theory.

e.  The traditional method of using a surface discharge on a glass or ceramic tube is clearly not appropriate as it depends on many atomic physics processes [3] and is too fragile for such a critical application. One possibility would be to use an inside-out washer gun configuration [3]. This would be a stack of alternating metal and insulating annular discs (washers) supported on an insulating sleeve surrounding the anode and providing high voltage isolation from the cathode. Gas could be injected from the small clearance between the insulating sleeve and washer stack so that it flows out through the gaps between the washers into space (or into the vacuum pump in a laboratory device). It should provide sufficient gas density for sequential breakdown of the gaps between the metal washers.

An interesting possibility, that could provide an independent motivation, would be to fill the hypodermic needle tube with deuterium gas at high pressure and look for neutron emission. The current front travelling over the outer surface should lead to a conical collapse of the hypodermic needle tube. This could form a cumulation jet [75,76] of a much higher velocity than the Alfven velocity estimated above. This would happen both because of increase in the velocity of the inner surface of the tube during convergence and because of the shaped charge effect [75,76]. The combination of high deuterium density, high confining magnetic field and possible incorporation of a poloidal bias magnetic field could make this a small-scale variant of the magnetized liner inertial fusion (MAGLIF) concept [18,19], with the cumulation jet playing the role of the laser igniter. This could be used to validate MAGLIF models. It might even be studied as a laboratory model for a potential aneutronic fusion interplanetary drive [77,78] for a Mars mission. Used without the deuterium, the hollow tube would still provide a higher velocity of ejection as compared to a solid wire because of the shaped charge effect.

The ballpark figures in Section IV show that a laboratory facility of a size similar to a UNU-ICTP plasma focus [3] should be sufficient to meet the objectives mentioned above as well as provide an early indication of the possibility of a fusion drive.

VI. <u>Summary and conclusion</u>:

This paper introduces and formulates a Generalized Plasma Focus problem that concerns a finite, axisymmetric plasma driven through a neutral gas medium at supersonic speed over distances much larger than its typical gradient scale length by its azimuthal magnetic field while remaining connected with its pulse power source through suitable boundaries. This results in a scaling theory, well-grounded in physics, that can calculate ballpark numbers for any newly conceived, experimentally untested configuration using a logical framework connected with first principles. Such scaling theory is necessary for the development of plasma-based space propulsion concepts based on modifications of the plasma focus device.

Predictions from this scaling theory are shown to compare well with the experimental knowledgebase.

This scaling theory is illustrated with a space propulsion concept that uses a modified plasma focus with a tapered anode. A metal wire (or hypodermic needle tube) extruded along the axis of the anode acts as a consumable portion of the anode that is rapidly vaporized by current transported behind the plasma sheath. It is shown that in a laboratory scale facility, a travelling wave with a magnetic field ~ 200 T traverses the surface of this wire in ~8 ns in an optimized configuration whose parameters are deduced from the scaling theory. When the wire is replaced with a hypodermic needle tube filled with deuterium, it may serve as an intense laboratory neutron source that could avoid the failure of neutron scaling in a conventional plasma focus. It could also be studied as a small-scale surrogate of the z-pinch dynamic hohlraum or MAGLIF concepts, as a laboratory model for an aneutronic fusion based interplanetary drive for a Mars mission or a non-nuclear electric propulsion engine for near-earth space missions.

Acknowledgement: This work was inspired by a conversation with my brother Arvind Auluck-Wilson.

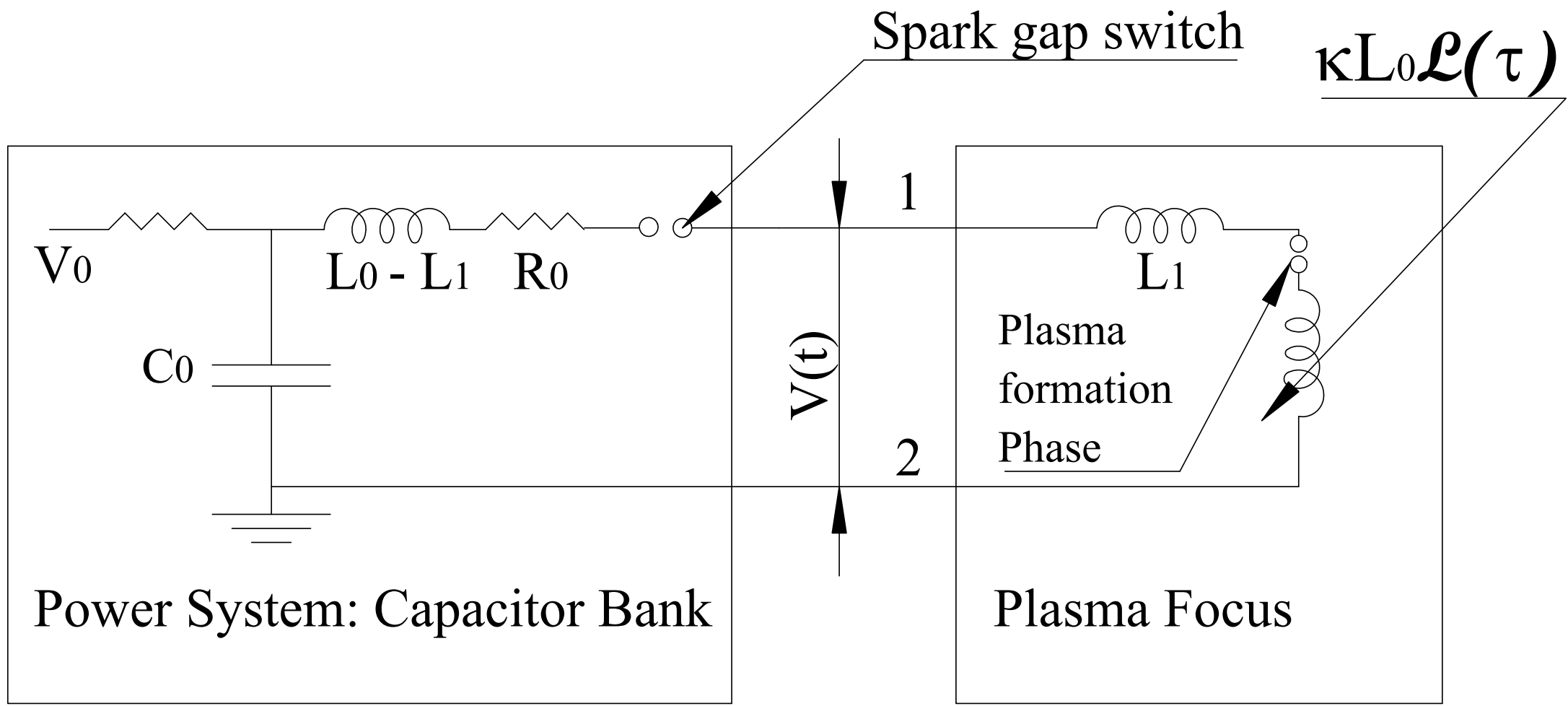

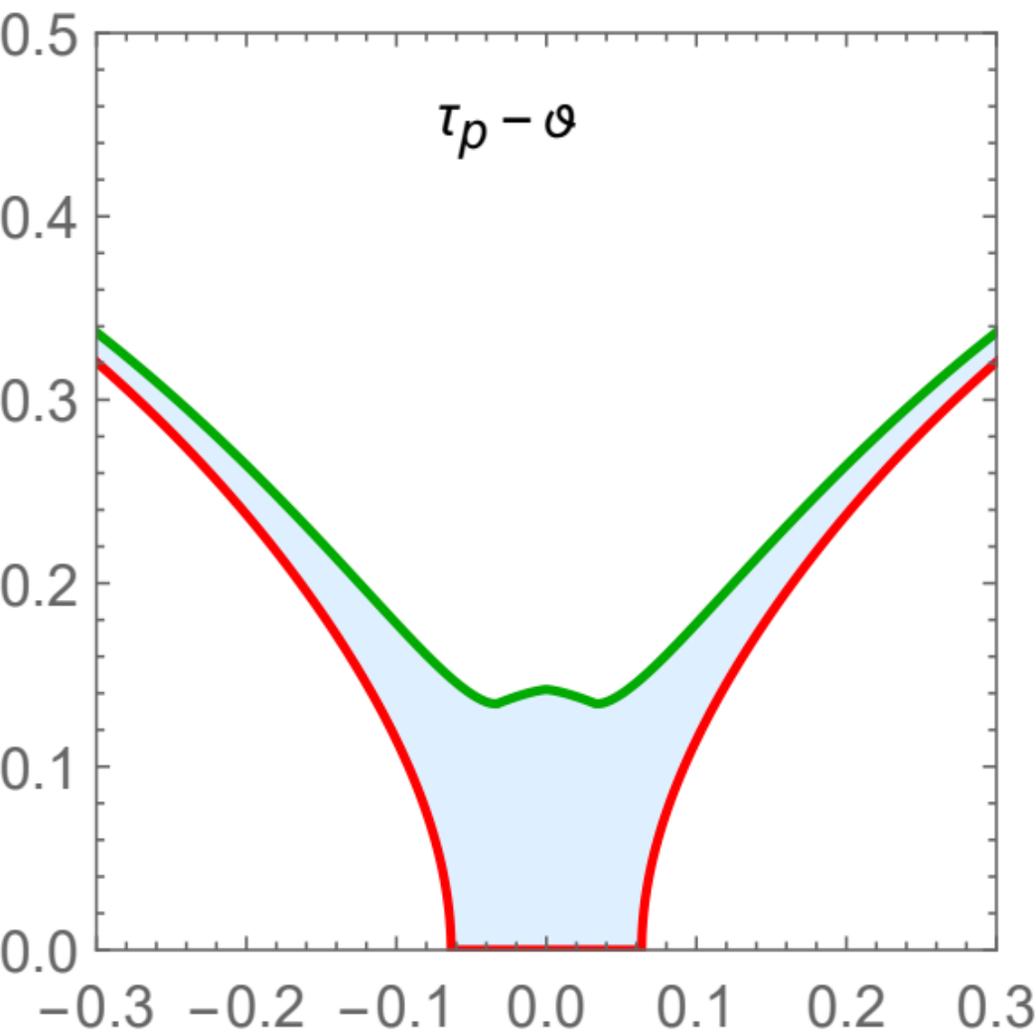

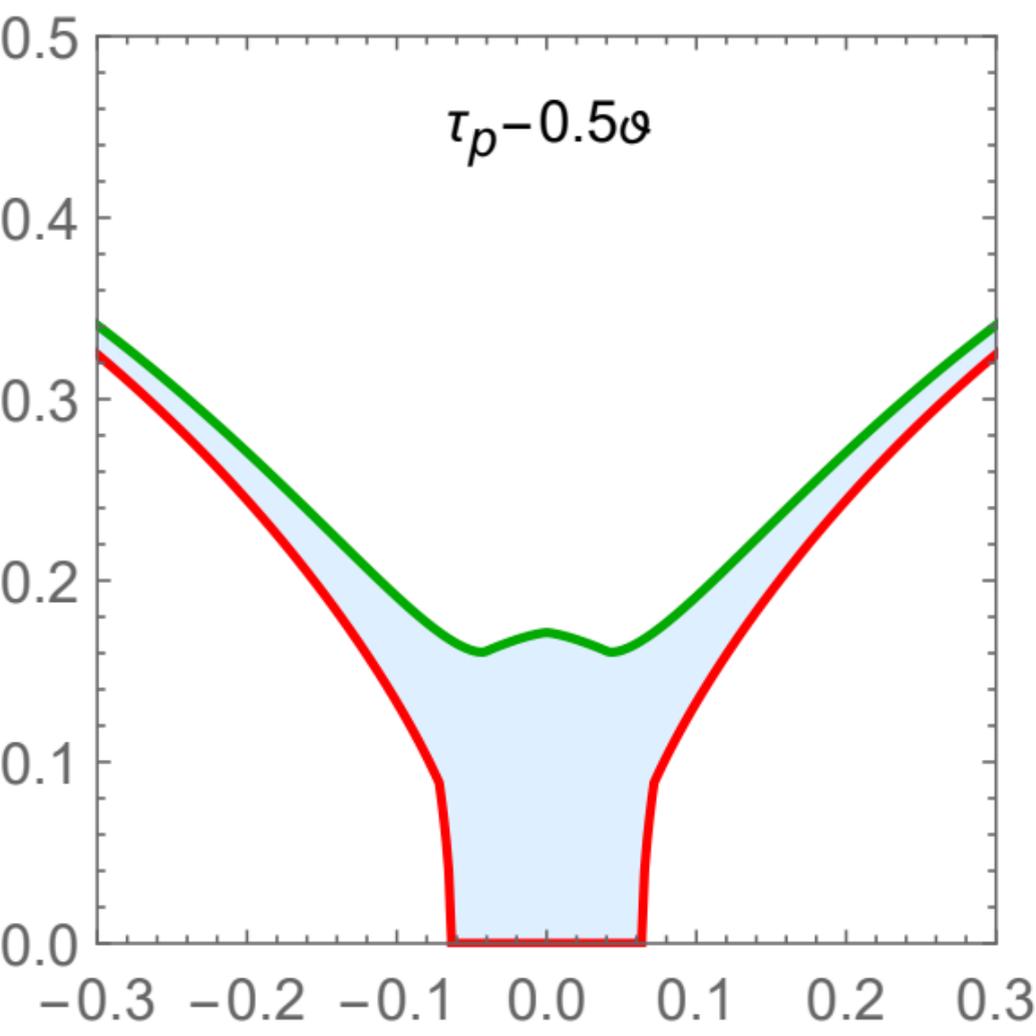

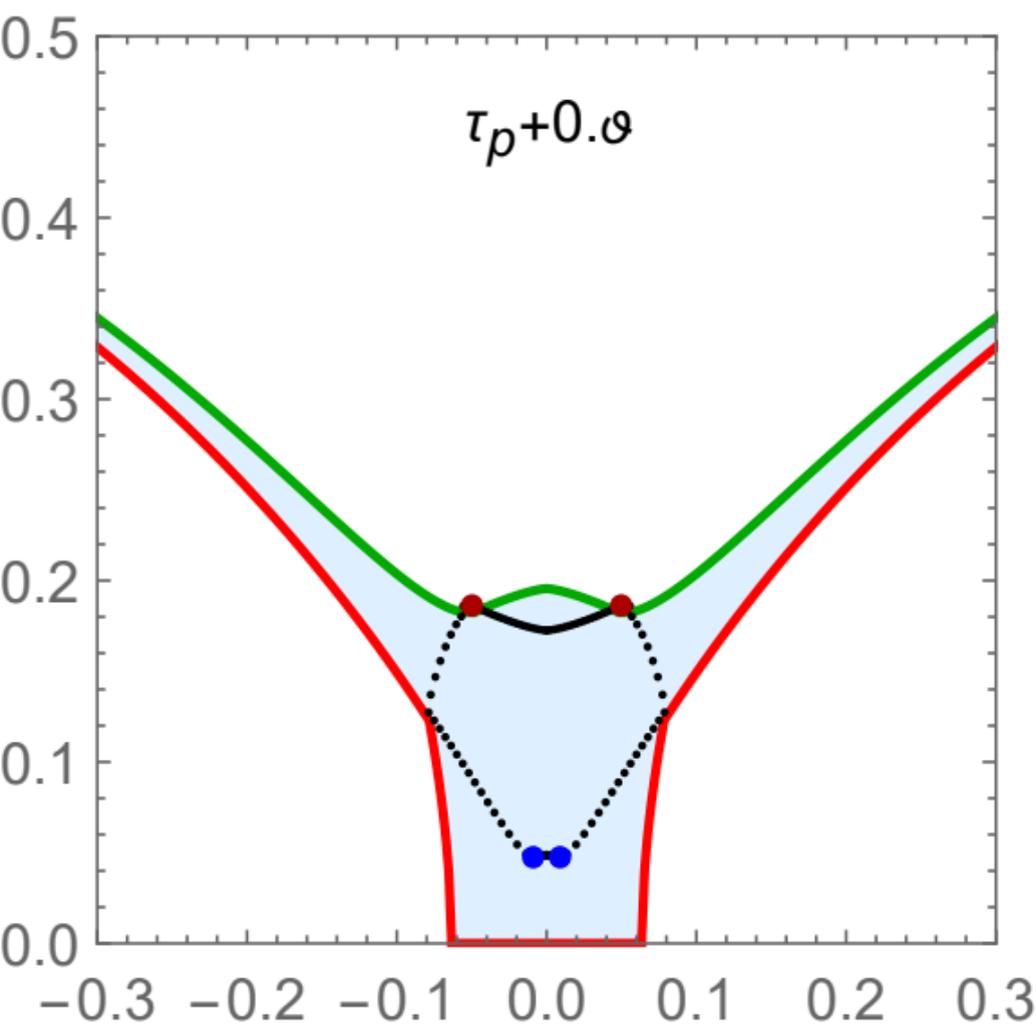

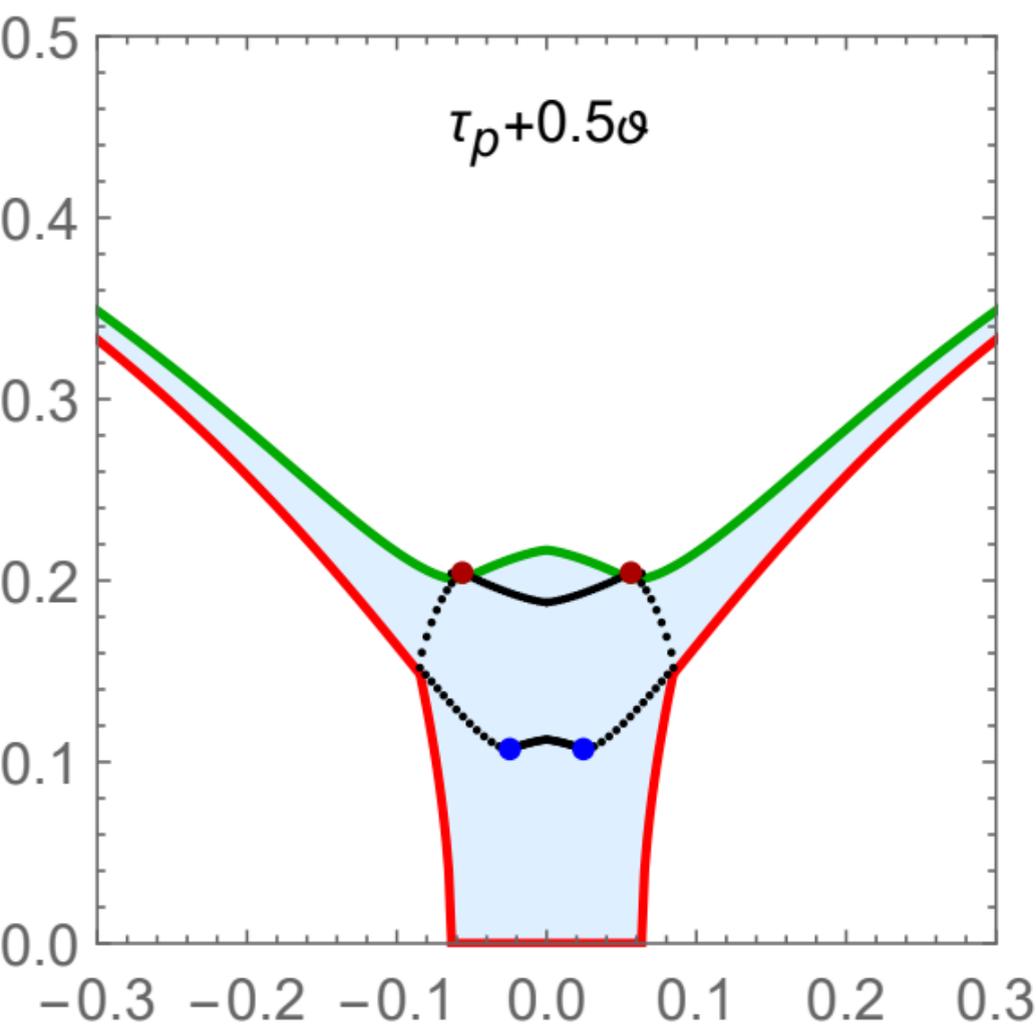

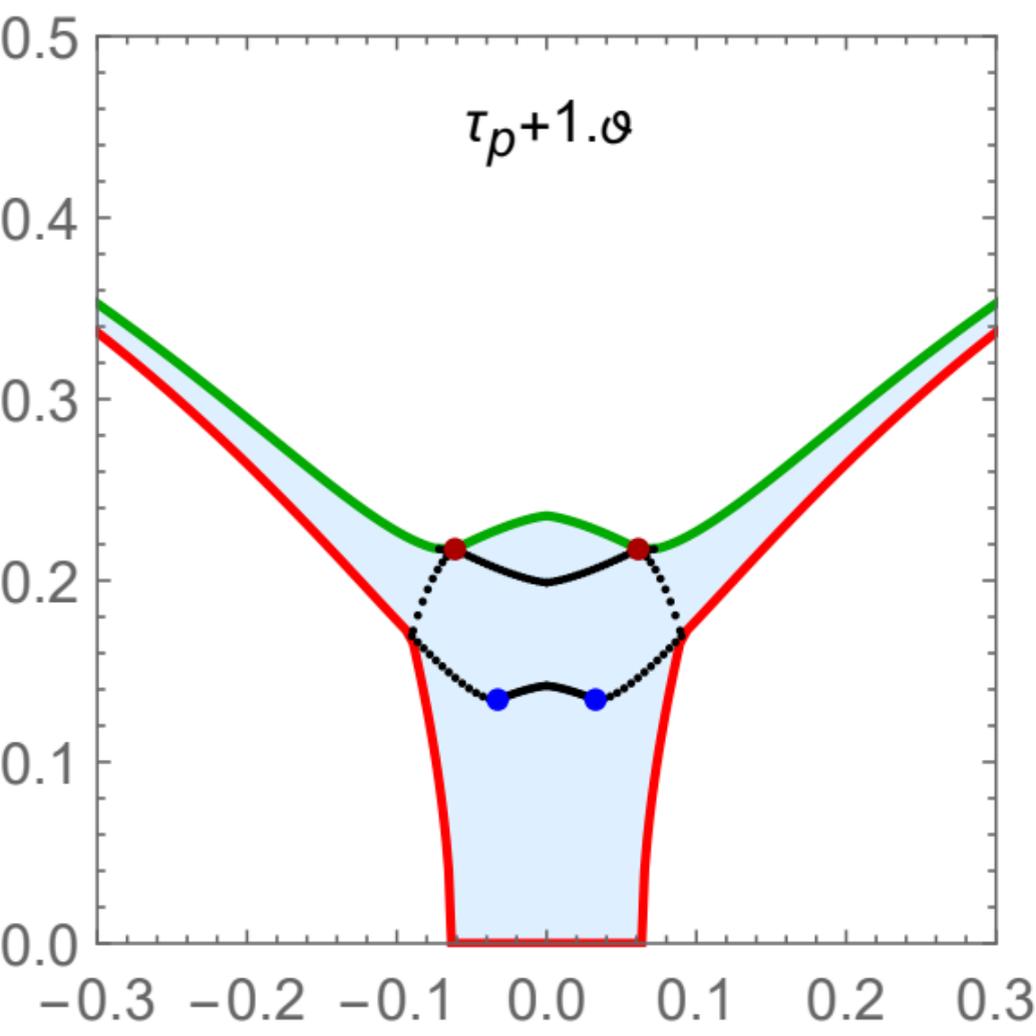

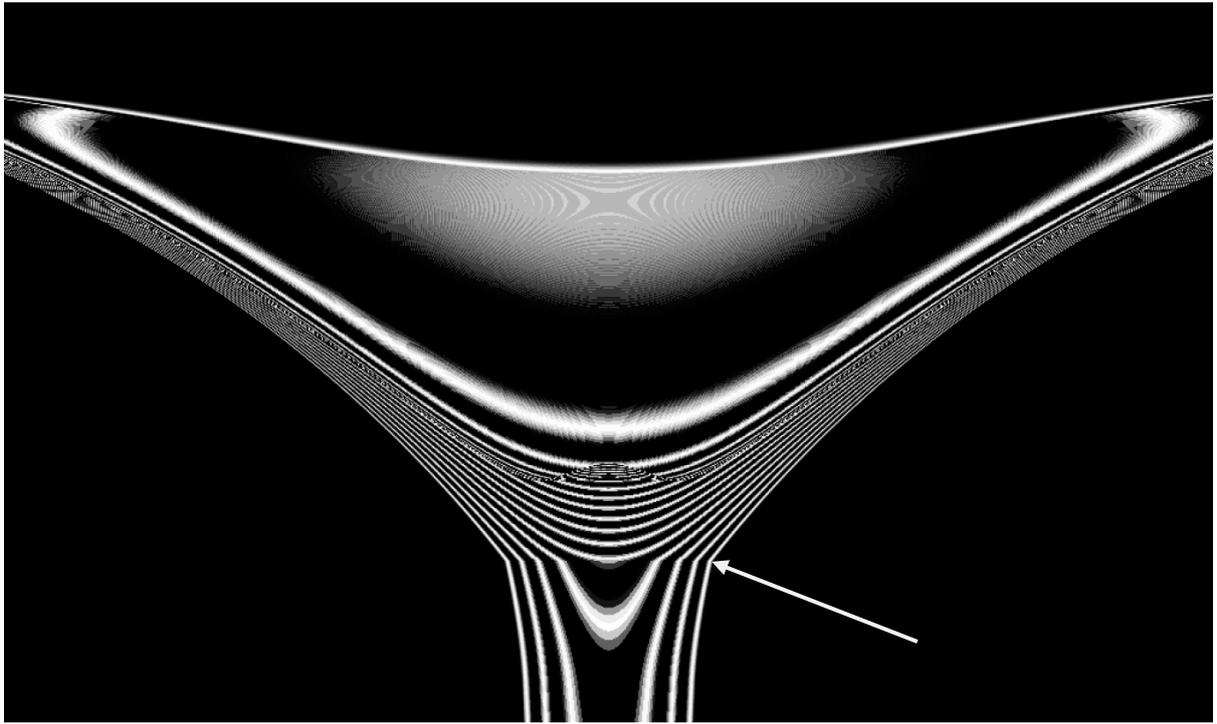

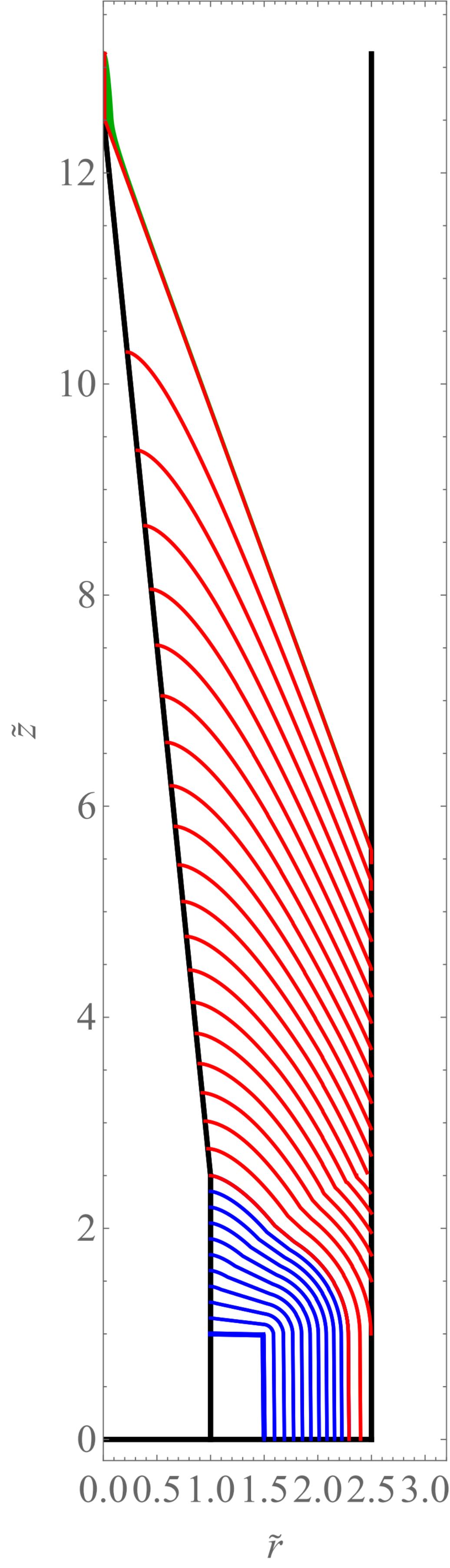

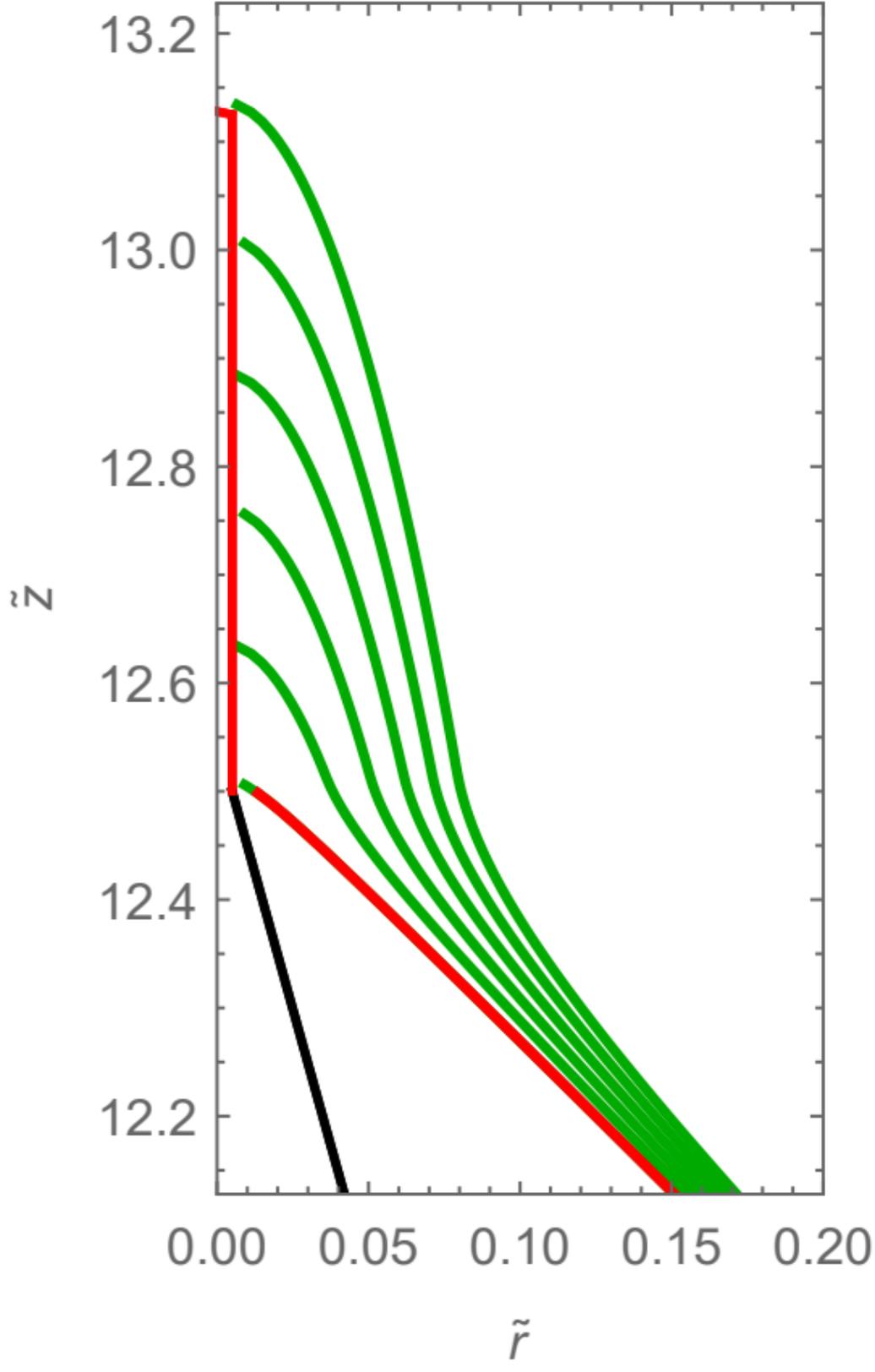

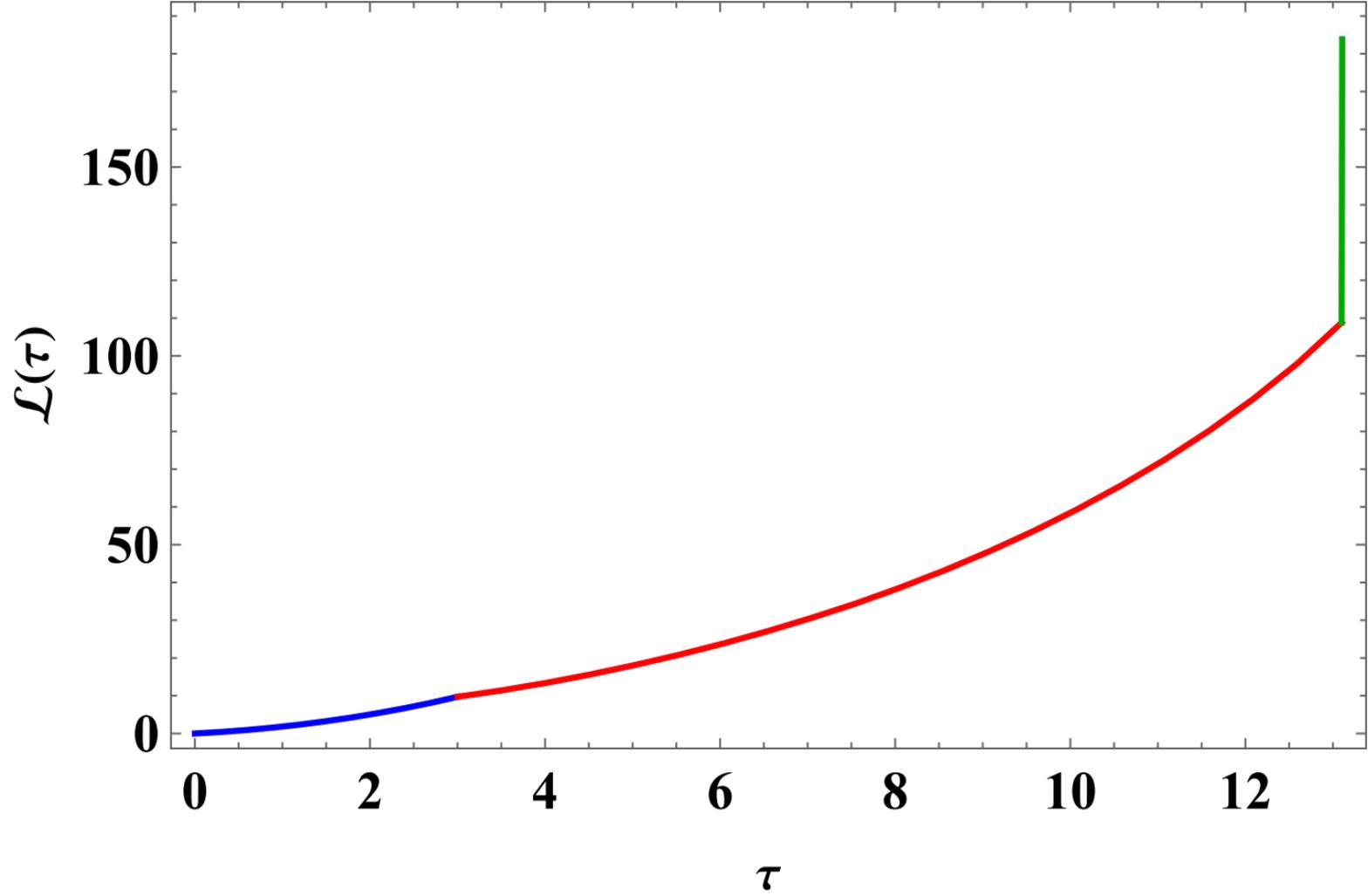

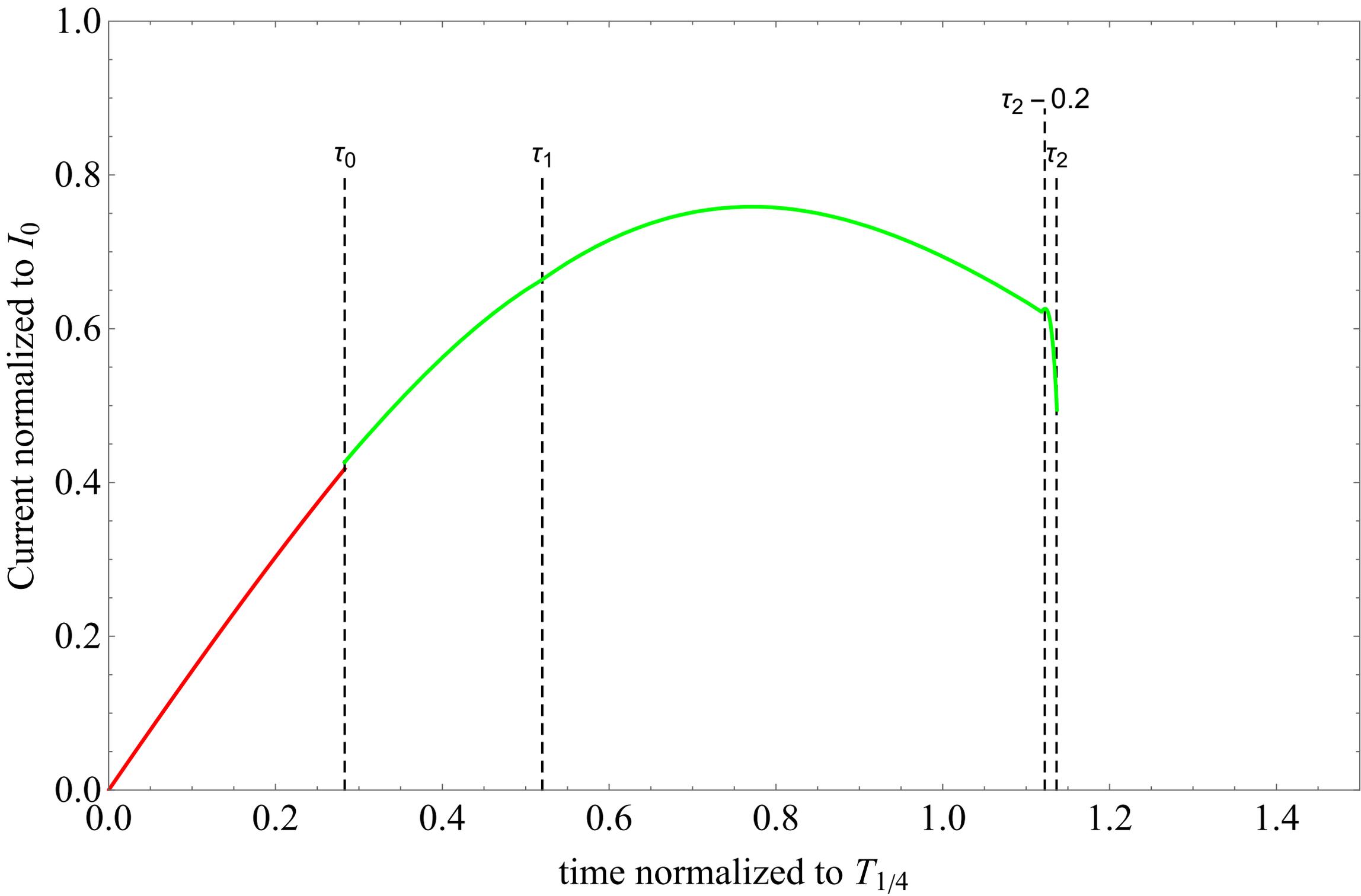

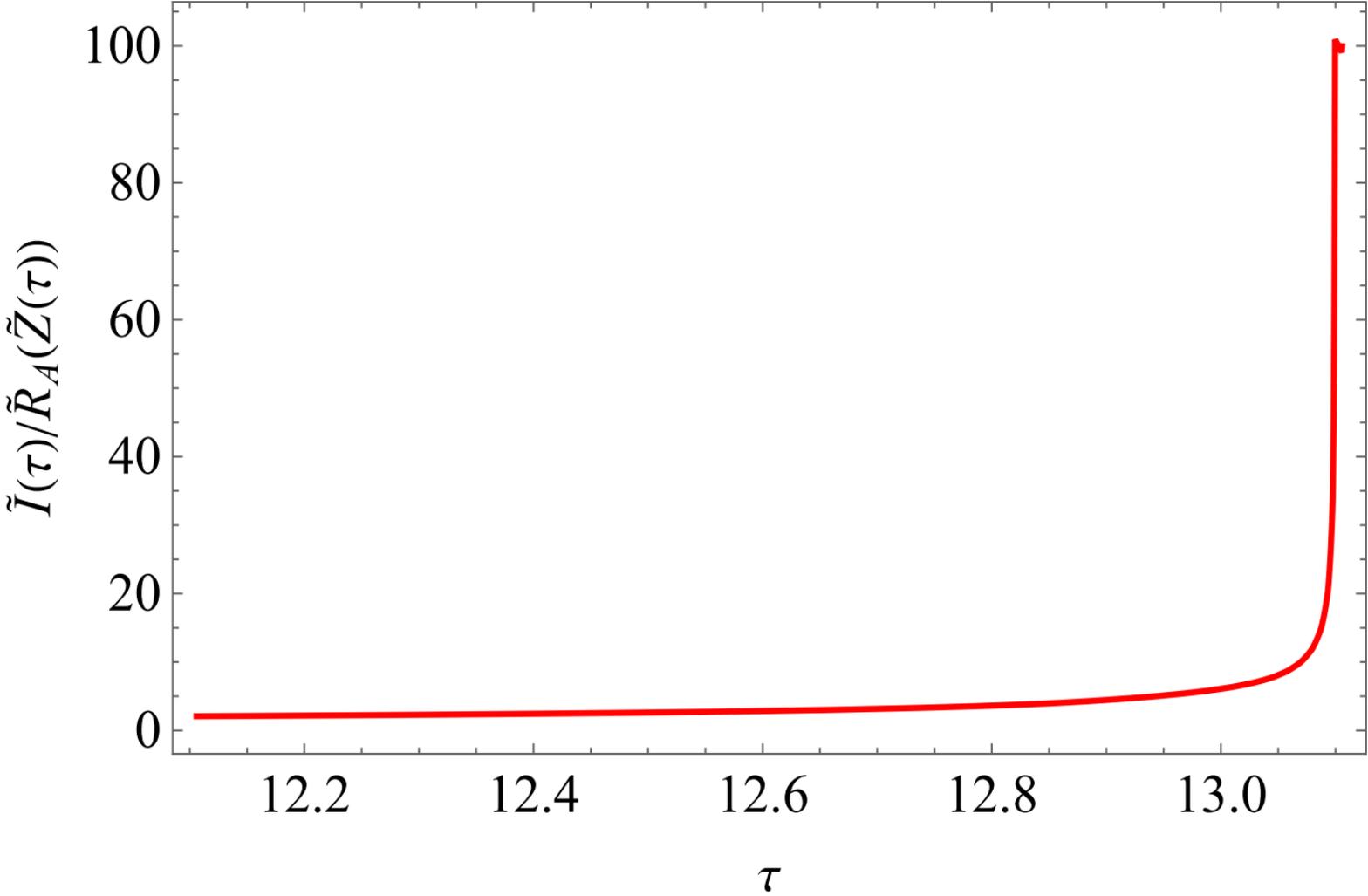

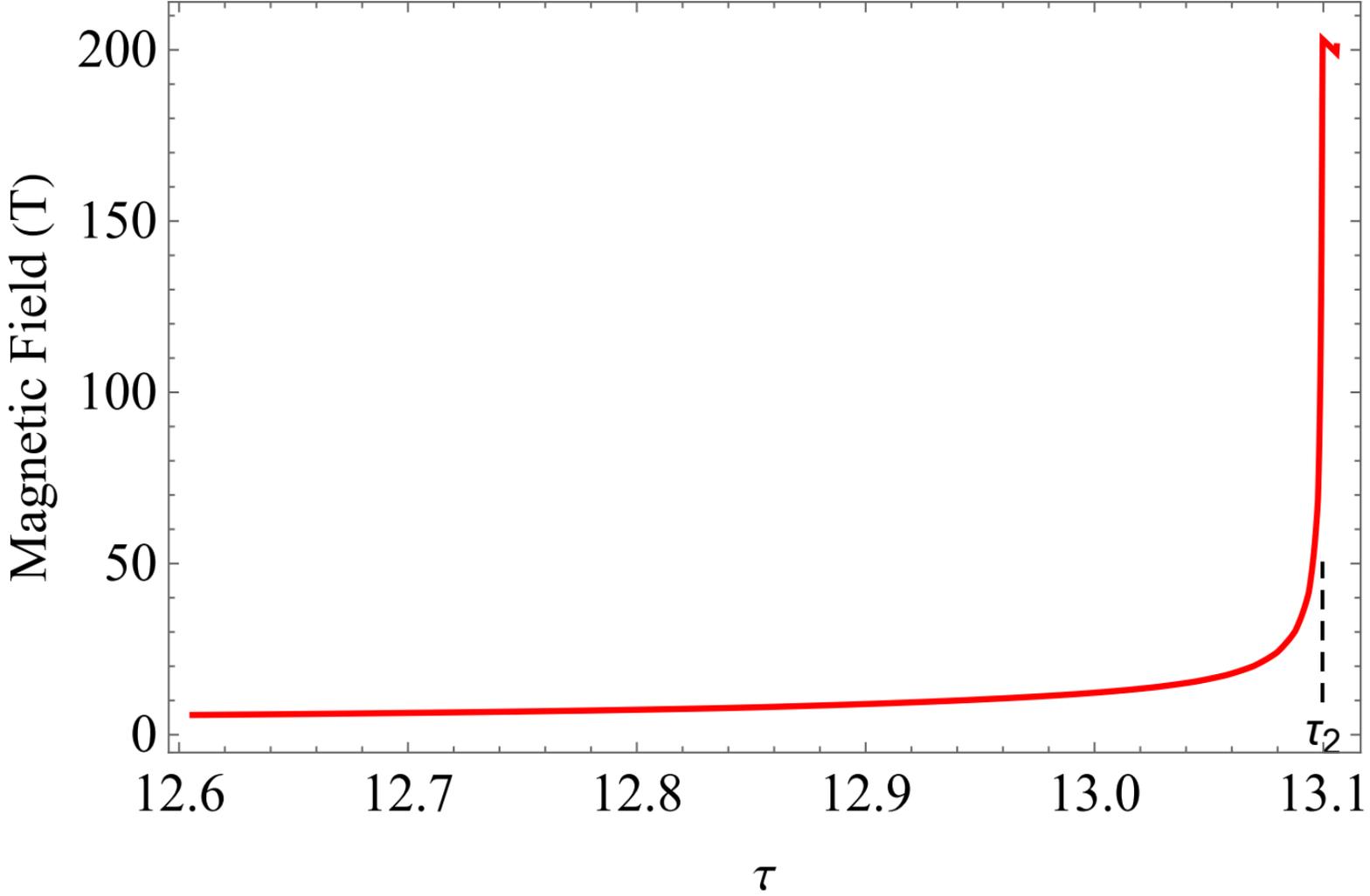